\documentclass[a4paper,USenglish]{article}

\usepackage[left=2.5cm,right=2.5cm,top=2.5cm,bottom=3cm]{geometry}
\usepackage[pdfborder={0 0 0}]{hyperref}
\usepackage{amsmath,amsthm,amssymb} 
\usepackage{thmtools,mathtools} 
\newtheorem{example}{Example}
\newtheorem{theorem}{Theorem}
\newtheorem{lemma}{Lemma}
\newtheorem{reductionrule}{Rule}
\newtheorem{observation}{Observation}
\newtheorem{corollary}{Corollary}
\newtheorem{openproblem}{Open Problem}
\newtheorem{fact}{Fact}
\usepackage{multirow}
\usepackage{booktabs}

\usepackage{caption}
\usepackage{enumitem}
\setlist[itemize]{noitemsep, topsep=1pt}
\setlist[enumerate]{noitemsep, topsep=1pt}

\newcommand\coloneq{\mathrel{\raise.4pt\hbox{:}{=}}}
\newcommand\eqcolon{\mathrel{{=}\raise.4pt\hbox{:}}}

\newcommand\Class[1]{%
  \mathchoice%
  {\text{\normalfont\small$\mathrm{#1}$}}%
  {\text{\normalfont\small$\mathrm{#1}$}}%
  {\text{\normalfont$\mathrm{#1}$}}%
  {\text{\normalfont$\mathrm{#1}$}}%
}
\newcommand{\Lang}[1]{\text{\normalfont\textsc{#1}}}

\DeclareMathOperator{\tw}{\mathrm{tw}}
\DeclareMathOperator{\poly}{\mathrm{poly}}
\DeclareMathOperator{\itw}{\mathrm{itw}}

\DeclareMathOperator{\cost}{\mathrm{cost}}
\DeclareMathOperator{\bag}{\chi}
\DeclareMathOperator{\children}{\mathrm{children}}
\DeclareMathOperator{\rootOf}{\mathrm{root}}

\def\phi{\varphi}
\DeclareMathOperator{\vars}{\mathrm{vars}}

\usepackage{tikz}
\usetikzlibrary{graphs, fit, backgrounds, positioning, arrows.meta, calc, bending}

\definecolor{jade}{rgb}{0.0, 0.66, 0.42}
\definecolor{cerise}{HTML}{CE4760}
\colorlet{fg}{jade!75!black}
\colorlet{bg}{cerise!75!black}

\newcommand{\ipu}[2][ipu]{%
  \foreach \x/\y in {#2}{
    \node[
      rounded corners,
      fill = gray,
      inner sep = 0pt,
      minimum width=1cm,
      minimum height=1cm,
    ] at (\x,\y) {};
    \node[
      rounded corners,
      fill = gray!60,
      inner sep = 0pt,
      minimum width=0.75cm,
      minimum height=0.75cm,
    ] at (\x,\y) {};
    \node[color=fg] at (\x,\y) {\textsc{ipu}};
    \draw[thick, color=fg, <-, >={[round]Stealth}] ($(\x,\y)+(0,0.15)$)     -- ++(0,0.5)  node (#1-input)  {};
    \draw[thick, color=fg, ->, >={[round]Stealth}] ($(\x,\y)+(0,-0.2)$)     -- ++(0,-0.5) node (#1-output) {};
    \draw[thick, color=fg, ->, >={[round]Stealth}] ($(\x,\y)+(-0.25,-0.2)$) -- ++(0,-0.5);
    \draw[thick, color=fg, ->, >={[round]Stealth}] ($(\x,\y)+(+0.25,-0.2)$) -- ++(0,-0.5);
  }
}

\title{Strong Structural Bounds for MaxSAT:\\The Fine Details of Using Neuromorphic and Quantum Hardware Accelerators}
\author{Max Bannach$^1$ \and Jai Grover$^1$ \and Markus Hecher$^2$}
\date{%
  $^1$European Space Agency, Noordwijk, 2201 AZ, The Netherlands \texttt{\{max.bannach,jai.grover\}@esa.int}\\[2ex]
  $^2$Massachusetts Institute of Technology, Computer Science and Artificial Intelligence Lab, USA \texttt{hecher@mit.edu}%
}

\begin{document}

\maketitle

\begin{abstract}
  Hardware accelerators like quantum annealers or
  neuromorphic chips are capable of finding the ground state of a
  Hamiltonian. A promising route in utilizing these devices is via
  methods from \emph{automated reasoning:} The problem at hand is first
  encoded into \Lang{maxsat}; then \Lang{maxsat} is reduced to
  \Lang{max2sat}; and finally,
  \Lang{max2sat} is translated into a Hamiltonian. It was
  observed that different encodings can dramatically
  affect the efficiency of the hardware accelerators. Yet, previous
  studies were only concerned with the size of the encodings rather
  than with syntactic or structural properties.
  
  We establish structure-aware reductions between \Lang{maxsat},
  \Lang{max2sat}, and the quadratic unconstrained binary optimization
  problem (\Lang{qubo}) that underlies such hardware accelerators. All
  these problems turn out to be equivalent under linear-time,
  treewidth-preserving reductions. As a consequence, we obtain tight
  lower bounds under $\Class{ETH}$ and $\Class{SETH}$ for
  \Lang{max2sat} and \Lang{qubo}, as well as a new time-optimal
  fixed-parameter algorithm for \Lang{qubo}. While our results are
  tight up to a constant \emph{additive} factor for the primal
  treewidth, we require a constant \emph{multiplicative} factor for
  the incidence treewidth. To close the emerging gap, we supplement
  our results with novel time-optimal algorithms for fragments of
  \Lang{maxsat} based on model counting.
\end{abstract}

\section{Introduction}

The \emph{maximum satisfiability} problem (\Lang{maxsat}) is the
canonical task of \emph{cost-optimal reasoning.} Given is a
\emph{weighted} pro\-positional formula in \emph{conjunctive normal
form} (\Lang{wcnf}) like
\begin{align*}
  \phi
  =\,
  &(x_1\vee x_2\vee x_3)^2
  \wedge
  (\neg x_1\vee x_2)^\infty
  \wedge
  (x_1\vee\neg x_2)^\infty\\
  \wedge\,
  &(\neg x_1\vee\neg x_2)^{5}
  \wedge
  (x_1\vee x_2\vee\neg x_3)^{120}
  \wedge
  (x_3)^ {100}.
\end{align*}
The goal is to find the model that \emph{minimizes the weights of the
falsified clauses.} (Here and in the rest of the article, the exponents
indicate the weights of the clauses.)  It is well-known in the theoretical
computer science literature that \Lang{maxsat} remains
$\Class{NP}$-hard if every clause contains at most two literals, i.e., 
\Lang{max2sat} is $\Class{NP}$-hard~\cite{GareyJS76}. This is in stark contrast to the
satisfiability problem, which is well known to be solvable in
polynomial time on such formulas~\cite{Krom67}. The hardness result of
\Lang{max2sat} imposes an encoding from \Lang{maxsat} into
\Lang{max2sat}, which, however, is so far deemed as of purely
theoretical interest. While there are well-established \Lang{maxsat}
solvers~\cite{MartinsML14,IgnatievMM19,Piotrow20} and specialized
solvers for fragments like Horn formulas~\cite{Marques-SilvaIM17}, we
are not aware of any tool that actively utilizes binary clauses.

The relevance of reductions from \Lang{maxsat} to
\Lang{max2sat} has increased during the last few years
due to rapid developments in specialized hardware
accelerators that are able to find the \emph{ground state} of an
\emph{Ising model.} These chips, called \emph{Ising machines}, \emph{Ising
processing units,} or \textsc{ipu}s~\cite{CoffrinNB19}, obtain as input a Hamiltonian
\[
H(x_1,\dots,x_n) = \sum_{i=1}^nw_ix_i + \sum_{i=1}^n\sum_{j=i+1}^n w_{ij}x_ix_j,
\]
with binary variables $x_i\in\{0,1\}$ (also called \emph{spins}) and weights
$w_{ij}\in\mathbb{Z}$.\footnote{The physics literature often
uses weights in $\mathbb{R}$, but any fixed-precision
approximation of $\mathbb{R}$ can be represented in $\mathbb{Z}$.} The ground state of the Ising model is the
vector $\vec x\in\{0,1\}^n$ that minimizes $H(\vec x)$. An
\textsc{ipu} is an abstraction of a chip that obtains as input just the
weights $w_{ij}$ (say in the form of an $n\times n$ upper triangular matrix) and that
outputs the ground state of the system,\footnote{Many realizations of \Lang{ipu}s,
such as the D-Wave quantum annealer, output only a heuristic
solution. In our theoretical model, however, we assume the \Lang{ipu}
to be an exact solver.} see Figure~\ref{figure:ipu}.

\begin{figure}[htbp]
  \centering
  \begin{minipage}{0.35\textwidth}
    \caption{Illustration of an \textsc{ipu} that operates on an
      $n$-variable Ising model $H(x_1,\dots,x_n)$. The input are the weights of $H$
      given as $n\times n$ upper triangular matrix ($n$ is fixed). The
      output is a vector $\vec x\in\{0,1\}^n$ that minimizes $H(\vec
      x)$.}
      \label{figure:ipu}
  \end{minipage}
  \begin{minipage}{0.64\textwidth}
    \hfill
    \raisebox{0.25cm}{\begin{tikzpicture}    
      \ipu{0/0}
      \node[above = 0.1cm of ipu-input]  {$H(x_1,\dots,x_n)$ as $W\in\mathbb{R}^{n\times n}$};
      \node[below = 0.1cm of ipu-output] {$x_1,\dots,x_n\in\{0,1\}^n$ minimizing $H(x_1,\dots,x_n)$};
    \end{tikzpicture}}
  \end{minipage}  
\end{figure}

The Ising model approach gained momentum since it naturally appeared
in various promising technologies simultaneously. Perhaps the
best-known \Lang{ipu} is the quantum annealer by
D-Wave~\cite{johnson2011quantum}, which claims to utilize quantum
effects to find the ground state. The usability of quantum-based
\Lang{ipu}s is, for instance, actively explored in satellite
scheduling~\cite{guillaume2022deep,AlHraishawiRC23} and for efficient
coalition formations among LEO satellites~\cite{VenkateshMNKD24}.
Another area in which Ising machines naturally appear are
\emph{neuromorphic chips}~\cite{AlomEMWT17a} such as Intel's
Loihi~\cite{DaviesWOSGJPR21} or \textsc{ibm}'s TrueNorth
system~\cite{AkopyanSCAAMIND15}.  This hardware implements physical
artificial neurons to perform computations, often in mixed
analog-digital implementations. It promises \emph{low-energy}
computations~\cite{SchumanKPMDK22} in \emph{radiation-resistant}
hardware~\cite{NaoukiIT23}, making it a promising technology for
future space applications~\cite{alves2024satellite,ortiz2024energy}.
The rising interest in Ising machines also led companies like Fujitsu
Limited~\cite{MatsubaraTMSWTT20} or Toshiba~\cite{KashimataYHT24} to
develop \Lang{ipu}s based on classical hardware.

The problem underlying the \Lang{ipu} (find the minimum of
$H\,{=}\,\sum_{i=1}^nw_ix_i+\sum_{i=1}^n\sum_{j=i+1}^n x_ix_jw_{ij}$) is called \emph{quadratic
unconstrained binary optimization} (\Lang{qubo}). Utilizing an
\Lang{ipu}, thus, requires to encode problems into
\Lang{qubo}s~\cite{Codognet23,codognet2024comparing}. A common route
in doing so is to use well-established encodings into \Lang{maxsat}
and to then encode \Lang{maxsat} into
\Lang{qubo}~\cite{BianCMRSV17,chancellor2016direct,MorseK23,ZielinskiBNLF24,NussleinZGLF23}. While
it was experimentally observed that the choice of the encoding can significantly impact the performance of
the Ising machine~\cite{KrugerM20,ZielinskiNSGLF23}, surprisingly, no
study has been performed so far that takes properties of the encoding into
account other than its pure size. In particular, to our
knowledge, there is no study that takes syntactic and structural
properties of the given \Lang{maxsat} 
instance into account~--~even though it is known that treewidth-based
approaches are \emph{competitive} for \Lang{maxsat}~\cite{BannachH24}
and that \Lang{ipu}s may \emph{only be usable} with Hamiltonians that
have small treewidth~\cite{wang2014ollivier}.

\subsection{\hspace{-.3cm}Contribution I: Treewidth-aware Reductions between MaxSAT and QUBO}

Our main results are fine-grained reductions between \Lang{maxsat}
variants and \Lang{qubo}. In fact, we prove that these problems are
equivalent under linear-time reductions that preserve the \emph{primal} or
\emph{incidence treewidth.} The primal treewidht $\tw(\phi)$ is the
treewidth of the graph $G_\phi$ that contains a vertex for every variable
and that connects variables that appear together in a clause; the
incidence treewidth $\itw(\phi)$ is the treewidth of the
variable-clause interaction graph $I_\phi$ that contains a vertex for
every variable and every clause and that connects variables to clauses
that contain them~--~formal definitions are given in Section~\ref{section:preliminaries:graphs}.
Figure~\ref{figure:overview} provides a more detailed overview of
Theorem~\ref{thm:main}.

\begin{theorem}[Main Theorem]\label{thm:main}
  Given a \Lang{wcnf} and a corresponding tree decomposition, the
  following reductions hold:
  \begin{enumerate}
  \item $\Lang{maxsat}\equiv^{\mathrm{lin}}_{\tw+2}\Lang{max2sat}\equiv^{\mathrm{lin}}_{\tw+2}\Lang{qubo}$
  \item $\Lang{maxsat}\equiv^{\mathrm{lin}}_{3\itw}\Lang{max2sat}\equiv^{\mathrm{lin}}_{3\itw}\Lang{qubo}$
  \end{enumerate}
  For Item~1, a tree decomposition is not required as input.
\end{theorem}

\begin{figure}[htbp]
  \centering
  \begin{minipage}{0.44\textwidth}
    \caption{An overview of our reductions between various variants of
      \Lang{maxsat}. All reductions in the picture are linear-time
      computable. An arrow \smash{$A\stackrel{a\tw+b}{\rightarrow} B$} means
      that if the instance of $A$ has primal treewidth $\tw$, the instance
      of $B$ has primal treewidth at most $a\tw+b$. If the
      label of an arrow is ``$\max(\tw,x)$'', the
      reduction increases the treewidth to at most
      $x$. Finally, if the arrow is dashed, the reduction requires a
      tree decomposition as input. The green arrows are implied by the transitive closure.}
    \label{figure:overview}
  \end{minipage}
  \begin{minipage}{0.54\textwidth}
    \raisebox{1cm}{\resizebox{\linewidth}{!}{\begin{tikzpicture}[
          reduction/.style = {
            semithick,
            ->,
            > = {[round,bend,sep]Stealth}
          }
        ]

        %
        % Problems
        %
        \node (maxsat)      at (0,0)  {\Lang{maxsat}};
        \node (max3sat)     at (3,0)  {\Lang{max3sat}};
        \node (max2sat)     at (6,0)  {\Lang{max2sat}};
        \node (mon-max2sat) at (6,-3) {\Lang{mon-max2sat}};
        \node (qubo)        at (0,-3) {\Lang{qubo}};

        %
        % Reduction arrows
        %
        \draw[reduction] (maxsat) to[bend left] node[midway, above]  {\scriptsize$\tw+2$} (max3sat);
        \draw[reduction, densely dashed] (maxsat) to[bend right] node[midway, below]  {\scriptsize$2\itw$} (max3sat);

        \draw[reduction] (max3sat) to[bend left] node[midway, above]   {\scriptsize$\max(\tw,3)$} (max2sat);
        \draw[reduction] (max3sat) to[bend right] node[midway, below]  {\scriptsize$3\itw$}       (max2sat);
        \draw[reduction, densely dashed] (maxsat) to[bend left=57.5] node[midway, below]  {\scriptsize$3\itw$} (max2sat);

        \draw[reduction] (max2sat) to[bend left] node[midway, right]  {\rotatebox{-90}{\scriptsize$\max(\tw,2)$}}  (mon-max2sat);
        \draw[reduction] (max2sat) to[bend right] node[midway, left]  {\rotatebox{-90}{\scriptsize$\max(\itw,2)$}} (mon-max2sat);

        \draw[reduction] (mon-max2sat) to[bend right=10] node[midway, above]  {\scriptsize$\tw$}  (qubo);
        \draw[reduction] (mon-max2sat) to[bend left=10]  node[midway, below]  {\scriptsize$\itw$} (qubo);

        \draw[reduction] (qubo) to[bend right] node[midway, right] {\rotatebox{-90}{\scriptsize$\max(\tw,2)$}}  (maxsat);
        \draw[reduction] (qubo) to[bend left]  node[midway, left]  {\rotatebox{-90}{\scriptsize$\max(\itw,2)$}} (maxsat);

        %
        % Transitive closure 
        %
        \draw[reduction, color=fg, densely dashed] (maxsat) -- ++(-2.5,0) -- node[midway, left]  {\rotatebox{-90}{\scriptsize$3\itw$}} (-2.5,-3) -- (qubo);
        \draw[reduction, color=fg] (maxsat) -- ++(-2,0)   -- node[midway, right]  {\rotatebox{-90}{\scriptsize$\tw+2$}} (-2,-3) -- (qubo);   
    \end{tikzpicture}}}      
  \end{minipage}
\end{figure}

\begin{corollary}[\Lang{qubo} Upper Bound]
  The ground state of a \Lang{qubo} $H$ can be computed in  $O\big(2^{\tw(H)}|H|\big)$.
\end{corollary}

\begin{corollary}[\Lang{max2sat} Lower Bound]\label{cor:maxlb}
  Unless $\Class{SETH}$ fails, solving $\Lang{max2sat}$ requires a
  running time of
  $\Omega(2^{\tw(\phi)})\cdot\poly(|\phi|)$ or $\Omega(2^{\itw(\phi)/3})\cdot\poly(|\phi|)$.
  
  Assuming $\Class{ETH}$, $\Lang{max2sat}$ cannot be solved in 
  $2^{o(\tw(\phi))}\cdot\poly(|\phi|)$ or $2^{o(\itw(\phi))}\cdot\poly(|\phi|)$.
\end{corollary}

\begin{corollary}[\Lang{qubo} Lower Bound]
  Unless $\Class{SETH}$ fails, $\Lang{qubo}$ requires a
  running time of
  $\Omega(2^{\tw(H)})\cdot\poly(|H|)$ or $\Omega(2^{\itw(H)/3})\cdot\poly(|H|)$.

  Assuming $\Class{ETH}$, $\Lang{qubo}$ cannot be solved in
  $2^{o(\tw(H))}\cdot\poly(|H|)$ or $2^{o(\itw(H))}\cdot\poly(|H|)$.
\end{corollary}

We refer the reader unfamiliar with the \emph{(strong) exponential
time hypothesis} ($\Class{SETH}$ and $\Class{ETH}$) to the
survey by Lokshtanov et al.~\cite{LokshtanovMS11}. Intuitively, in the context of
structural parameters, $\Class{ETH}$ implies that \Lang{3sat} cannot be solved in
$2^{o(\tw(\phi))}\poly(|\phi|)$ while $\Class{SETH}$ implies \Lang{sat} cannot be solved in
$o(2^{\tw(\phi)})\poly(|\phi|)$.

\subsection{Contribution II: New Algorithms for Variants of MaxSAT}

The reductions in Figure~\ref{figure:overview} imply new
lower bounds for \Lang{qubo} and \Lang{max2sat} that match the
corresponding upper bounds for the primal treewidth. For the incidence
treewidth, the implied lower bounds have the form
$\Omega(2^{\itw(\phi)/3}\poly(|\phi|))$ while the implied upper bounds
have the form $O(2^{2\itw(\phi)}\poly(|\phi|))$ (via the folklore
dynamic program). We narrow
this gap by improving the upper bound:

\begin{theorem}\label{theorem:binaryitw}
  \Lang{max2sat} and \Lang{qubo} can be solved in time $O(2^{\itw(\phi)}|\phi|)$.
\end{theorem}

The central part of the proof of the theorem uses the fact that
clauses and terms in both problems are \emph{binary.} We supplement these
findings with new upper bounds for fragments of \Lang{maxsat} that
have \emph{unbounded clause sizes.} The idea is to establish
treewidth-aware reductions from these problems to \Lang{\#sat}, the
problem of \emph{counting} the number of satisfying assignments of an
(unweighted) propositional formula.

%{\color{red}\textbf{TODO: update}
\begin{theorem}\label{thm:max2sat}~\hfill\\[-0.5cm]\raggedright
  \begin{enumerate}
  \item $\Lang{unary-maxsat}\leq^{\mathrm{lin}}_{\itw+1}\Lang{\#sat}$
  \item $\Lang{mult-maxsat}\leq^{\mathrm{lin}}_{\itw+1}\Lang{\#sat}$
  \item $\Lang{lex-maxsat}\leq^{\mathrm{lin}}_{\itw+1}\Lang{\#sat}$
\end{enumerate}
\end{theorem}
%\begin{proof}
%  
%\end{proof}

\begin{corollary}\label{cor:wcnf}
  \Lang{unary-maxsat}, \Lang{mult-maxsat}, and \Lang{lex-maxsat} can
  be solved in time $O\big(2^{\itw(\phi)}|\phi|\big)$.
\end{corollary}
\begin{proof}
Use Theorem~\ref{thm:max2sat} and an
algorithm with this running time for \Lang{\#sat}~\cite{SlivovskySzeider20}.
\end{proof}

\subsection{Related Work}
A reduction from \Lang{sat} to \Lang{max2sat} is known since
1976~\cite{GareyJS76}. This reduction was improved by Trevisan et
al.~\cite{TrevisanSSW00} and is still a field of research.  An
overview of recent developments is given by Ansotegui et
al.~\cite{AnsoteguiL21}. Connections between \Lang{maxsat} and it's
multiplicative and lexicographic version, i.e., \Lang{mult-maxsat} and
\Lang{lex-maxsat}, are well known~\cite{Marques-SilvaAGL11}.

Reducing \Lang{maxsat} to its binary version turned out to be a
central part of embeddings into adiabatic quantum computers, quantum
annealers~\cite{abs-2403-00182,BianCMRSV17,chancellor2016direct,DatePSP19,RodriguezFarresBALC24,Santra2014},
and systems that emulate artificial intelligence in hardware such as
neuromorphic chips~\cite{AlomEMWT17a,Mniszewski19}. Consequently,
optimal (with respect to some property) reductions from \Lang{maxsat}
to \Lang{max2sat} and \Lang{qubo} are actively
researched~\cite{MorseK23}. Current trends include algorithmic
formulations~\cite{NussleinGLF22} and data-driven automatic generation
of \Lang{qubo}s~\cite{MoraglioGS22}. However, to the best of our
knowledge, no work considered structural properties of these
reductions so far.

\subsection{Structure of this Article}

The following section provides the necessary background in
propositional logic and structural graph theory. The subsequent
section proves the main theorem as a series of \emph{reduction rules.}
Each of these rules obtains as input a formula of some fragment of
\Lang{maxsat} and modifies it slightly. The proof of
Theorem~\ref{thm:main} boils down to showing that applying these rules
exhaustively results in a \Lang{max2sat} or \Lang{qubo} instance, and
that these rules can be applied exhaustively in linear time without
increasing the input's treewidth. We extend this result with new upper
bounds for fragments of \Lang{maxsat} using tools from probabilistic
reasoning and conclude in the last section. 

\section{Preliminaries}\label{section:preliminaries}

A \emph{literal} $\ell$ is variable $x$ or its negation $\neg x$. We
refer to the variable of a literal with $|\ell|$. A \emph{clause} is a \emph{set of literals}, and a
propositional formula in \emph{conjunctive normal form} (a \Lang{cnf})
is a \emph{set of clauses}. A \emph{weighted} \Lang{cnf} (a
\Lang{wcnf}) is a \Lang{cnf} $\phi$ together with a \emph{weight
function} $w\colon\phi\rightarrow\mathbb{N}\cup\{\infty\}$. Clauses
with finite weight are called \emph{soft,} and the others \emph{hard}.
We denote formulas in the usual logical notation and use
the superscript of clauses to denote their weight. For instance, the
formula from the introduction refers to
\begin{align*}
\phi \equiv \big\{
\{x_1, x_2, x_3\},
\{\neg x_1, x_2\},
\{x_1,\neg x_2\},
\{\neg x_1,\neg x_2\},
\{x_1, x_2,\neg x_3\},
\{x_3\}
\big\}
\end{align*}
with, e.g., $w(\{x_1, x_2, x_3\})=2$.
We denote the variables in $\phi$ with $\vars(\phi)$. An
\emph{assignment}~$\beta$ of $\phi$ is a set of literals with
$\{\,|\ell|\mid\ell\in\beta\,\}=\vars(\phi)$ and
$|\beta|=|\vars(\phi)|$. An assignment $\beta$ satisfies a clause $c$
if $\beta\cap c\neq\emptyset$, which we denote by $\beta\models
c$. The \emph{cost} of an assignment is the sum of the falsified
clauses:
\(
\cost(\beta) = \sum_{c\in\phi,\beta\not\models c}w(c).
\)
The cost of a formula $\phi$ is the minimum cost any assignment~$\beta$
of $\phi$ must have, i.e., $\cost(\phi)=\min_\beta\cost(\beta)$. An assignment
\emph{satisfies} a formula ($\beta\models\phi$) if it has a finite cost.
\begin{example}
  For the previous formula we have $\cost(\phi)\leq 5$ witnessed by $\cost(\{x_1,x_2,x_5\})=5$ (only
  $(\neg x_1\vee\neg x_2)$ is falsified). It is easy to
  check that indeed $\cost(\phi)=5$.
\end{example}

We define the monotone version of \Lang{maxsat} as the problem in which \emph{all} variables
occur \emph{negated}. Since this problem is trivial if all clauses
have a non-negative weight, we allow
$w\colon\phi\rightarrow\mathbb{Z}\cup\{\infty\}$ for monotone formulas.

\subsection{Structural Graph Theory for Propositional Logic.}\label{section:preliminaries:graphs}
A graph $G$ consists of a set of \emph{vertices} $V(G)$ and a set $E(G)$ of
two-element subsets of $V(G)$ called \emph{edges}. The \emph{primal
graph} $G_\phi$ of a \Lang{wcnf} $\phi$ is the graph with
$V(G_\phi)=\vars(\phi)$ that connects two vertices if the
corresponding variables appear together in a clause. The
\emph{incidence graph} $I_\phi$ is the bipartite graph with
$V(I_\phi)=\phi\cup\vars(\phi)$ that connects variables to the clauses
containing them.

A \emph{tree decomposition} of a graph $G$ is a (rooted) directed tree $T$ with
a mapping $\bag\colon V(T)\rightarrow 2^{V(G)}$ such that:
\begin{description}
\item[Connectedness] For every $v\in V(G)$ the induced subgraph
  $T[\{t\in V(T)\mid v\in\bag(t)\}]$ is a nonempty directed
  tree.
\item[Covering] For every $\{u,v\}\in E(G)$ there is a $t\in V(T)$
  with $\{u,v\}\subseteq\bag(t)$.    
\end{description}
The \emph{width} of $(T,\bag)$ is $\max_{t\in V(T)}|\bag(t)|-1$, and
the treewidth $\tw(G)$ of $G$ is the minimum width of any tree
decomposition of $G$. The \emph{primal and incidence treewidth} of a
formula $\phi$ is the treewidth of $G_\phi$ and $I_\phi$, respectively:
\[
\tw(\phi)\coloneq \tw(G_\phi)
\quad\text{and}\quad
\itw(\phi)\coloneq \tw(I_\phi).
\]
We do not enforce any further requirements on tree decompositions, but
note that since $T$ is rooted, there is a unique $\rootOf(T)$ and every
node $t\in V(T)$ has a set $\children(t)$. For a vertex $v\in V(G)$ we
denote with $\rootOf(T,v)$ the root of the subtree $T[\{t\in V(T)\mid v\in\bag(t)\}]$.

\begin{example}
  The treewidth of the Centaurus constellation (as graph shown on the
  left) is at most two, as proven by the tree decomposition on the
  right:
  
  \null\hfill
  \scalebox{0.6}
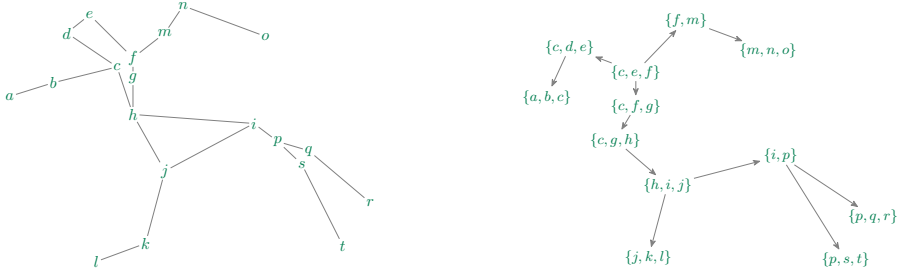
{\begin{tikzpicture}[
      star/.style = {
        color     = fg,
        inner sep = 1pt
      }
    ]
    \node[star] (a) at (-3.75,0.9)    {$a$};
    \node[star] (b) at (-2.8,1.2)     {$b$};
    \node[star] (c) at (-1.4,1.55)    {$c$};
    \node[star] (d) at (-2.5,2.3)     {$d$};
    \node[star] (e) at (-2,2.7)       {$e$};
    \node[star] (f) at (-1.05,1.75)   {$f$};
    \node[star] (g) at (-1.05,1.3)    {$g$};
    \node[star] (h) at (-1.05,.5)     {$h$};
    \node[star] (i) at (1.6,.3)       {$i$};
    \node[star] (j) at (-0.35,-.75)   {$j$};
    \node[star] (k) at (-0.775,-2.35) {$k$};
    \node[star] (l) at (-1.85,-2.75)  {$l$};
    \node[star] (m) at (-0.35,2.3)    {$m$};
    \node[star] (n) at (0.05,2.9)     {$n$};
    \node[star] (o) at (1.85,2.225)   {$o$};
    \node[star] (p) at (2.125,-.1)    {$p$};
    \node[star] (q) at (2.8,-.3)      {$q$};
    \node[star] (r) at (4.15,-1.45)   {$r$};
    \node[star] (s) at (2.65,-0.6)    {$s$};
    \node[star] (t) at (3.55,-2.4)    {$t$};
    
    \graph[use existing nodes, edges = {semithick, color=gray}]{
      a -- b -- c -- d -- e -- f -- g -- h -- i -- j -- h -- c;
      j -- k -- l;
      i -- p -- q -- r;
      p -- s -- t;
      f -- m -- n -- o;
    };
  \end{tikzpicture}}
  \hfill
  \scalebox{0.6}{\begin{tikzpicture}[
      bag/.style = {
        color     = fg,
        inner sep = 1pt,
        font      = \small
      }
    ]
    \node[bag] (b1) at (-3,1.2)       {$\{a,b,c\}$};
    \node[bag] (b2) at (-2.5,2.3)     {$\{c,d,e\}$};
    \node[bag] (b3) at (-1.05,1.75)   {$\{c,e,f\}$};
    \node[bag] (b4) at (0.05,2.9)     {$\{f,m\}$};
    \node[bag] (b5) at (1.85,2.225)   {$\{m,n,o\}$};
    \node[bag] (b6) at (-1.05,1)      {$\{c,f,g\}$};
    \node[bag] (b7) at (-1.5,.25)     {$\{c,g,h\}$};
    \node[bag] (b8) at (-0.35,-.75)   {$\{h,i,j\}$};
    \node[bag] (b9) at (-0.775,-2.35) {$\{j,k,l\}$};
    \node[bag] (b10) at (2.125,-.1)   {$\{i,p\}$};
    \node[bag] (b11) at (4.15,-1.45)  {$\{p,q,r\}$};
    \node[bag] (b12) at (3.55,-2.4)   {$\{p,s,t\}$};
    
    \graph[use existing nodes, edges = {semithick, color=gray, >={[round,sep]Stealth}}]{
      b1 <- b2 <- b3 -> b4 -> b5;
      b3 -> b6 -> b7 -> b8 -> {b9, b10};
      b10 -> {b11, b12};
    };
  \end{tikzpicture}}
  \hfill\null%
\end{example}

All these definitions extend to \emph{quadratic binary optimization
problems} \Lang{qubo}s in a natural way: Just interpret 
a Hamiltonian $H\,{=}\,\sum_{i=1}^nw_ix_i+\sum_{i=1}^n\sum_{j=i+1}^n x_ix_jw_{ij}$ as a
set of sets, where each of the inner sets corresponds to a term. 

\section{From MaxSAT to QUBO and Back}\label{section:reductions}

All statements in this section are proven with a series of
\emph{reduction rules.} Each rule obtains as input a \Lang{wcnf}
$\phi$ and is either \emph{applicable}, in which case a new $\phi'$ is
output, or not, in which case nothing happens. When applying a
rule~$i$, we always assume that all rules $j<i$ are not applicable. In
any reduction rule, $\alpha$ will refer to a fresh auxiliary variable
\emph{produced by the application of the rule}. To illustrate this idea, the
following two rules transform any \Lang{wcnf} into the standard form
in which all soft clauses are non-trivial and unit.

\begin{reductionrule}\label{rule:trivial}
  Applicable if there is a $c=(\ell_1\vee\dots\vee\ell_q)^w$ with
  $w=0$.

  Delete $c$.
\end{reductionrule}

\begin{reductionrule}\label{rule:unitsoft}
  Applicable for $c=(\ell_1\vee\dots\vee\ell_q)^w$ with
  $w<\infty$ and $q>1$.
  
  Replace $c$ with
  $(\alpha)^w\wedge(\ell_1\vee\dots\vee\ell_q\vee\neg \alpha)^\infty$.
\end{reductionrule}

\begin{lemma}\label{lemma:trivial}
  Rules~\ref{rule:trivial} and~\ref{rule:unitsoft} can be
  applied exhaustively in linear time, increase the primal treewidth
  by at most one, and do not increase the incidence treewidth.
\end{lemma}
\begin{proof}
  The first rule can be applied at most $|\phi|$ times and takes, assuming
  any reasonable data structure to store $\phi$, time $O(1)$. For the
  second rule, we just have to scan the clauses once and apply a
  constant number of operations per clause if necessary.

  Since treewidth is hereditary, Rule~\ref{rule:trivial} cannot
  increase it. To see that Rule~\ref{rule:unitsoft} does increase
  the primal treewidth by at most one, take a bag of a tree
  decomposition of~$G_\phi$ that contains all variables of $c$. Add
  a new bag that contains all variables of $c$ and~$\alpha$.

  For the incidence case, there is a bag containing~$c$, to which we
  can attach a bag $\{c,\alpha\}$. Note that if
  Rule~\ref{rule:unitsoft} is applicable, there must be a bag
  containing at least two elements~--~hence,
  the incidence treewidth is not increased.
\end{proof}

\subsection{From MaxSAT to Max3SAT}

The first step of our reductions is to ensure that every clause
contains at most three literals. While this is a well-known operation,
we must ensure that the treewidth is not
increased. This turns out to be rather easy for the primal treewidth
and quite tricky for the incidence treewidth.

\begin{reductionrule}\label{rule:max3sat}
Applicable if there is a 
$c=(\ell_1\vee\dots\vee\ell_q)^\infty$ with
$q>3$.

Replace $c$ with
$(\ell_1\vee\ell_2\vee\alpha)^\infty\wedge\allowbreak(\neg\alpha\vee\ell_{3}\vee\dots\vee\ell_{q})^\infty$.      
\end{reductionrule}

\begin{lemma}\label{lemma:w3cnfptw}
  Rule~\ref{rule:max3sat} can be applied exhaustively in linear time and
  increases the primal treewidth by at most one.
\end{lemma}
\begin{proof}
  The rule can be applied at most $|c|-3$ times per clause
  and, thus, overall at most $O(\sum_{c\in\phi}|c|)$ times~--~which is
  linear in the size of the formula.

  For the claim on the treewidth, take a bag that contains all
  variables of $c$. Attach a new bag $b$ containing the variables
  of~$c$ and $\alpha$ (this may increase the treewidth by one). Attach
  to this bag two new ones, one containing
  $\{|\ell_1|,|\ell_2|,\alpha\}$ and one containing
  $\{|\ell_3|,\dots,|\ell_q|,\alpha\}$. In repeated
  applications of the rule, we never have to add a variable to $b$: we can always
  create a new copy of the original bag or work with the two smaller
  bags~--~hence, the treewidth is increased by at most one in total.
\end{proof}

Unfortunately, it is unclear in how far Rule~\ref{rule:max3sat}
affects the incidence treewidth. The clause $c$ and its variables may
be ``smeared'' across an optimal tree decomposition. Therefore, arbitrarily
splitting the clause may increase the treewidth in an uncontrolled way. However, if
we have access to a tree decomposition, we can apply the rule in a
more controlled way to avoid this effect. 
Lampis et al.~\cite[Proposition A.1]{LampisMM18}~showed how to apply the rule exhaustively such that the
incidence treewidth grows from $k$ to $O(k)$ (using a complex program
that processes the given tree decomposition). The following lemma
improves this result in two ways: It is a direct encoding that does
not require an involved program, and it determines the hidden factor to be two:

\begin{lemma}\label{lemma:itw3cnf}
  There is a linear-time algorithm that, given a \Lang{wcnf}~$\phi$ and a
  width-$k$ tree decomposition of $I_\phi$, outputs a \Lang{w3cnf}
  $\psi$ with $\itw(\psi)\leq2k$  and $\cost(\phi)=\cost(\psi)$. 
\end{lemma}
\begin{proof}
  We prove the claim in three steps: \emph{(i)} we present the
  encoding, then \emph{(ii)} we show its correctness, and finally
  \emph{(iii)} we establish the claim about the
  treewidth. Importantly and in contrast to previous reduction rules,
  here we have \emph{access to a tree decomposition} to \emph{guide} our encoding. 

  For part \emph{(i)}, let $\phi$ be the given \Lang{wcnf} and
  $(T,\bag)$ be a width-$k$ tree decomposition of the incidence graph
  $I_\phi$. Our encoding consists of two parts: \emph{distributed
  variables} that allow access to the variables of $\phi$ in a
  controlled way along the tree decomposition; and an \emph{evaluation
  of clauses via simulated dynamic programming} that ensures the
  semantics of $\phi$. Let $\psi$ be the formula that we construct, and
  let
  \[
  \vars(\psi) =
  \bigcup_{t\in V(T)}\qquad\, \bigcup_{\mathclap{x\in\vars(\phi)\cap\bag(t)}}\,\,\{x^t\}
  \,\,\cup
  \bigcup_{t\in V(T)}\quad \bigcup_{\mathclap{c\in\phi\cap\bag(t)}}\,\,\{c^t\}.
  \]

  \paragraph*{Distributed Variables:} The first part of our
  encoding ensures that for all $x\in\vars(\phi)$, all the copies $x^t$
  in $\psi$ always have the same value. We will also say that these
  copies are \emph{synchronized.} To that end, we introduce for every $t\in
  V(T)$, every $x\in \vars(\phi)\cap\bag(t)$, and for every
  $t'\in\{t'\in\children(t)\mid x\in\bag(t')\}$ the following clauses:
  \begin{align}\label{constraint:sync}
  (x^t\leftrightarrow x^{t'})^\infty
  \equiv
  (\neg x^t\vee x^{t'})^\infty
  \wedge
  (x^t\vee \neg x^{t'})^\infty.
  \end{align}
  \paragraph*{Simulated Dynamic
    Programming:} To encode the semantics of $\phi$ to $\psi$, we need
  to enforce that the variables corresponding to clauses are set to true
  at some point. This is achieved by adding for every $c\in\phi$ the
  following clause:
  \begin{align}\label{constraint:cost}
    (c^{\rootOf(T,c)})^{w(c)}.
  \end{align}
  This (eventually soft) constraint says that the copy of $c$ closest to the root should
  be true. The weight thereby equals the weight of $c$. To conclude
  the encoding, whenever a $c^t$ is set to true, there must be a
  \emph{reason:} either one of the literals of $c$ in $\bag(t)$ is
  true, or $c$ was already satisfied in a child of $t$. The following
  clause realizes this semantic:
  \begin{align}\label{constraint:clauses}
    \big(
    c^t\rightarrow\qquad
    \bigvee_{\mathclap{\substack{\ell\in c\\|\ell|\in\bag(t)
      \\|\ell|\not\in\cup_{t'\in\children(t)}\bag(t')      
    }}}\,\,\ell
    \qquad\vee\qquad
    \bigvee_{\mathclap{\substack{t'\in \children(t)\\c\in\bag(t')}}}\,\,c^{t'}\,\,\,
    \big)^\infty.
  \end{align}

  This concludes the description of the encoding. To 
  establish \emph{(ii)}, we first show
  $\cost(\psi)\leq\cost(\phi)$ by proving that for every assignment
  $\alpha\models\phi$ there is an assignment $\beta\models\psi$ with
  $\cost(\alpha)=\cost(\beta)$. Given $\alpha$, we first define a
  \emph{partial} assignment that satisfies~(\ref{constraint:sync}):
  \[
  \beta' \coloneq\quad
  \bigcup_{\mathclap{\substack{t\in V(T)\\x\in\vars(\phi)\cap\bag(t)}}}
  \,\,\{x^t\mid x\in\alpha\}
  \cup
  \{\neg x^t\mid \neg x\in\alpha\}.
  \]
  We extend $\beta'$ to $\beta''$ by adding for every clause
  $c\in\phi$ with $\alpha\models c$ the literal $c^{\rootOf(T,c)}$ to
  $\beta'$. By~(\ref{constraint:cost}), we have
  $\cost(\alpha)=\cost(\beta'')$. We claim $\beta''$ can be extended
  to a satisfying assignment $\beta\models\psi$, for which we need to
  establish~(\ref{constraint:clauses}). Since $\alpha\models c$, there
  is a literal $\ell\in c\cap\alpha$. By the definition of $I_\phi$
  and the second property of tree decompositions,
  there is a bag $t\in V(T)$ with $\{|\ell|,c\}\subseteq\bag(t)$,
  i.e., $\{c^t,|\ell|^t\}\subseteq\vars(\psi)$. According to the
  polarity of $\ell$, we add $|\ell|^t$ or $\neg|\ell|^t$ to
  $\beta''$. Then, adding $c^t$ to $\beta''$ is consistent
  with~(\ref{constraint:clauses}) at~$t$. We obtain $\beta$ by adding
  $c^{t'}$ for every $t'\in V(T)$ that lies on the unique path from
  $t$ to $\rootOf(T,c)$, and~$\neg c^{t''}$ for all other $t''\in V(T)$ with
  $c\in\bag(t'')$.
  
  The other direction, i.e., $\cost(\phi)\leq\cost(\psi)$, works
  similar: For every assignment $\beta\models\psi$ we construct
  an assignment $\alpha$ with $\alpha\models\phi$:
  \[
  \alpha \coloneq
  \bigcup_{x\in\vars(\phi)}
  \begin{cases}
    \phantom{\neg}x^t & \text{if there is a $x^t$ in $\beta$;}\\
    \neg x^t & \text{else.}
  \end{cases}
  \]

  Symmetric to the first direction, we have $\alpha\models c$ for all
  $c\in\phi$ with $\beta\models c^{\rootOf(T,c)}$ (but $\alpha$ may also
  satisfy clauses with $\beta\not\models c^{\rootOf(T,c)}$). Hence, we
  have $\cost(\alpha)\leq\cost(\beta)$.

  We continue with~\emph{(iii)} and prove that $\itw(\psi)\leq 2k$
  if the given tree decomposition of $I_\phi$ has width~$k$. To that
  end, we take the given tree decomposition $(T,\bag)$ of $I_\phi$ and
  morph it into one of $I_\psi$. First, we rename in every bag $t\in
  V(T)$ nodes $x\in\vars(\phi)\cap\bag(t)$ to $x^t$ and
  $c\in\phi\cap\bag(t)$ to $c^t$. To every bag that contains a
  $c^{\rootOf(T,c)}$, we attach a new bag containing
  $c^{\rootOf(T,c)}$ and~(\ref{constraint:cost}). 

  To continue, we add to $\bag(t)$ the
  constraint~(\ref{constraint:clauses}) for every $c\in\phi\cap\bag(t)$,
  which may double the size of~$\bag(t)$. This establishes the
  properties of an incidence tree decomposition for the first part
  of~(\ref{constraint:clauses}), e.g., the big-or over the
  literals. To establish the second part, e.g., the big-or over the
  children, we need to add the constraint to all children of
  $t$. This can be achieved without increasing the width 
  by adding a chain of bags between $t$ and the child $t'$. The first
  element of the chain contains $\bag(t)\setminus\{c^{t}\}\cup \{c^{t'}\}$; then the
  next bag does not need to contain $c^{t'}$, but may contain a
  $\tilde c^{t'}$ for another clause~$\tilde c$. This process is
  repeated until we reach~$t'$, to which nothing further is added. The
  bags along this chain have the size of~$\bag(t)$, e.g.,
  $2k$. Finally, to establish~(\ref{constraint:sync}), we proceed
  similarly: While connecting $t$ and $t'$, we one-by-one add and 
  forget vertices $x^{t'}$. Then, on the chain from $t$ to $t'$ there is a
  bag containing $x^{t}$ and $x^{t'}$, to which we can attach two new
  bags containing these variables and one of the clauses
  of~(\ref{constraint:sync}).

  This almost concludes the proof: We have established a $\psi$ with
  $\cost(\phi)=\cost(\psi)$, which has incidence treewidth at most
  $2k$. However, $\psi$ is not a \Lang{w3cnf} due
  to~(\ref{constraint:clauses})~--~and building such a formula is the
  whole point of the lemma! Fortunately, this can easily be fixed: It
  is well known that any tree decomposition can be transformed in
  linear time into one in which every node has either exactly one
  child introducing at most one new vertex, or two children with the
  same content~\cite[Chapter~7]{CyganFKLMPPS15}. Transforming
  $(T,\bag)$ to this special form before applying the encoding from
  above ensures that~(\ref{constraint:clauses}) contains at most three
  literals. 
\end{proof}

\subsection{From Max3SAT to Max2SAT}

To reduce from \Lang{max3sat} to \Lang{max2sat}, we first remove hard
clauses by replacing them with soft clauses of high weight (referred
to as \emph{top value}).
\begin{reductionrule}\label{rule:allsoft}
  Applicable if $\phi$ contains hard clauses. Define
  \[
  h\coloneq 1+\sum_{\mathclap{c\in\phi, w(c)<\infty}}w(c)
  \quad\text{and}\quad
  \theta\coloneq\sum_{\mathclap{c\in\phi,w(c)=\infty}}h
  \]
  and change the weight of every hard clause to $h$. 
\end{reductionrule}

Let $\phi'$ be the formula obtained from $\phi$ by applying
Rule~\ref{rule:allsoft}.
Then it is easy to see and well known that $\phi$ is satisfiable iff
$\cost(\phi')<\theta$ and that in that case $\cost(\phi)=\cost(\phi')$. It is also clear that
$\tw(\phi)=\tw(\phi')$ and $\itw(\phi)=\itw(\phi')$ as the rule only modifies
weights. To move from \Lang{max3sat} (without hard clauses) to
\Lang{max2sat}, we use:

\begin{reductionrule}\label{rule:max2sat}
  Applicable if there is a 
  $c=(\ell_1\vee\ell_2\vee\ell_3)^h$. Replace~$c$ with 
  $(\ell_1\vee\ell_2)^h$, $(\ell_1\vee\ell_3)^h$,
  $(\neg\ell_2\vee\neg\ell_3)^h$, $(\alpha\vee\neg\ell_1)^h$,
  $(\neg\alpha\vee\ell_2)^h$, and
  $(\neg\alpha\vee\ell_3)^h$. Increase $\theta$ by $4h$.
\end{reductionrule}

It was proven by Ansotegui et al.~\cite[Lemma~3]{AnsoteguiL21}
that Rule~\ref{rule:max2sat} is correct (observe that the set of clauses is unsatisfiable and that an
assignment that satisfies $c$ satisfies exactly 5 of them, while an
assignment violating $c$ satisfies only 4 of the new clauses).

\begin{lemma}\label{lemma:max2sat}
  Rule~\ref{rule:max2sat} can be applied exhaustively in linear time and 
  increase the primal treewidth by at most one and the incidence
  treewidth by at most a factor of three.
\end{lemma}
\begin{proof}
  To apply the rule exhaustively, scanning over the formula once and
  printing a constant number of constant size clauses per encountered
  clause is sufficient.  This is possible in time $O(|\phi|)$.

  For the statement about the primal treewidth, observe that there must be a
  bag $t$ containing $|\ell_1|$, $|\ell_2|$, and $|\ell_3|$. We
  attach a new bag $t'$ with
  \(
  \bag(t')=\{|\ell_1|,|\ell_2|,|\ell_3|,\alpha\}
  \)
  and obtain a tree decomposition of $\phi'$.

  For the incidence treewidth, the situation is slightly more
  complex. In a tree decomposition of $I_\phi$ there are three bags $t_1$,
  $t_2$, and $t_3$ with $\{c,|\ell_i|\}\subseteq\bag(t_i)$ for
  $i\in\{1,2,3\}$, but there is no guarantee that any of the three
  variables appear together in a bag. By the properties of a tree
  decomposition, however, $c$ is contained in any bag that lies on a
  path between two of these bags. Hence, we can substitute $c$ in all
  bags by $\alpha,|\ell_1|,|\ell_2|$ without violating a property
  of the tree decomposition. Then
  $\{\alpha,|\ell_1|,|\ell_2|,|\ell_3|\}\subseteq\bag(t_3)$. This
  increases the treewidth by at most a factor of three and allows us
  to attach new bags to $t_3$ that connect the new clauses to the
  variables they contain.
\end{proof}

\begin{corollary}
  Rule~\ref{rule:max2sat} increases the primal treewidth to at most $\max(\tw,3)$.
\end{corollary}

It is worth pointing out a difference to the previous section: While
Rule~\ref{rule:max2sat} also increases the incidence treewidth by a
multiplicative factor, it does \emph{not} need a tree decomposition as
input. If we are given a tree decomposition, we can extend
Lemma~\ref{lemma:itw3cnf} and directly output a \Lang{w2cnf}:

\begin{lemma}\label{lemma:itw2cnf}
  There is a linear-time algorithm that, given a \Lang{wcnf}~$\phi$ and a
  width-$k$ tree decomposition of $I_\phi$, outputs a \Lang{w2cnf}
  $\psi$ with $\itw(\psi)\leq 3k$ and $\cost(\phi)=\cost(\psi)$. 
\end{lemma}
\begin{proof}
  Use the encoding of Lemma~\ref{lemma:itw3cnf} and utilize
  Rule~\ref{rule:max2sat} for constraint~(\ref{constraint:clauses}). The
  properties of a tree decomposition can be restored by simply adding
  $\alpha$ to all bags containing~(\ref{constraint:clauses}), which
  increases the treewidth to $3k$.
\end{proof}

\subsection{From Max2SAT to Monotone Max2SAT}

In order to encode a \Lang{max2sat} formula $\phi$ into a
\Lang{qubo}, we first transform $\phi$ such that every clause has the
form $(\neg x)$ or $(\neg x\vee\neg y)$, i.e., contains one or two negative literals. Such
formulas are called \emph{monotone}; hence, we reduce
\Lang{max2sat} to \Lang{mon-max2sat}. Since every literal is negated
in this problem, it represents a restriction of \Lang{horn-max2sat}.

Recall that after applying the previous
rules, all clauses are soft, and a top value $h$ that simulates hard
clauses is defined. A simple way to achieve the form is the use of
double-railed logic by replacing every variable $x$ with $x^+$ and
$x^-$, every occurrence of $x$ with $\neg x^-$, and every occurrence
of $\neg x$ with $\neg x^+$. We then encode $(x^+\leftrightarrow
\neg x^-)^h$ via $(\neg x^+\vee\neg x^-)^{2h}$, $(\neg x^+)^{-h}$, and
$(\neg x^-)^{-h}$. Observe that, in contrast to the definition of
\Lang{maxsat}, we do allow negative weights here~--~otherwise
\Lang{mon-maxsat} is a trivial problem. Also observe how the sum of the
weights of the three clauses is $h$ if, and only if, $(\neg
x^+\vee\neg x^-)$ evaluates to true. Unfortunately, this simple reduction increases
even the primal treewidth by a factor of two. To preserve the
treewidth as claimed, we perform the above sketched reduction
\emph{locally}.

\begin{reductionrule}\label{rule:monotone}
  Applicable if there is a clause $c=(x\vee\ell)^w$ with a positive
  literal $x$ (and an arbitrary literal $\ell$). Replace $c$ by 
  $(\neg\alpha\vee \ell)^w$,
  $(\neg\alpha\vee\neg x)^{2h}$,
  $(\neg\alpha)^{-h}$, and
  $(\neg x)^{-h}$.
\end{reductionrule}

\begin{lemma}\label{lemma:monotone}
  Rule~\ref{rule:monotone} can be applied exhaustively in linear time
  and increases both the primal and incidence treewidth by at most one.
\end{lemma}
\begin{proof}
  The claim about the running time follows once more by the simple
  observation that every clause is replaced by a constant number of
  new clauses.

  To see that the primal treewidth increases by at most one,
  take any tree decomposition of $G_\phi$. There is a bag containing
  $x$ and $|\ell|$, to which we can attach a new bag containing $x$,
  $|\ell|$, and $\alpha.$

  For the incidence treewidth, consider a
  tree decomposition of $I_\phi$. We first replace $c$ in every bag with
  $\alpha$. Then there is a bag containing $\alpha$ and $|\ell|$
  to which we attach a new bag containing $\alpha$, $|\ell|$, and
  $(\neg\alpha\vee\ell)$. Similarly, there are bags to which we can
  attach the other new clauses without increasing the width further.
\end{proof}

\begin{corollary}
  Rule~\ref{rule:monotone} increases the primal treewidth to
  at most $\max(\tw,2)$ and the incidence treewidth to at most $\max(\itw,2)$.
\end{corollary}

\subsection{From Monotone Max2SAT to QUBO}

Once we have a monotone \Lang{wcnf}, it is straightforward to
transform it into a \Lang{qubo}:

\begin{reductionrule}\label{rule:monotone2qubo}
  Applicable if there is a clause $c=(\neg x\vee\neg y)^w$. Replace $c$ by 
  the term $-wxy$.
\end{reductionrule}

\begin{observation}
  Rule~\ref{rule:monotone2qubo} can be applied exhaustively in linear
  time and neither increases the primal nor the incidence treewidth.
\end{observation}

The output of Rule~\ref{rule:monotone2qubo} is a Hamiltonian $H$ whose
ground states correspond to optimal models of the input formula. We
collect the results that we have established in this section as:

\begin{theorem}\label{theorem:maxsat2qubo}
  Applying Rules~\ref{rule:trivial}--\ref{rule:monotone2qubo}
  exhaustively is possible in linear time and establishes a reduction
  from \Lang{maxsat} to \Lang{qubo} that increases the primal
  treewidth by at most one and the incidence treewidth by at most a
  factor of three.
\end{theorem}

\subsection{From QUBO to MaxSAT}

Encoding a \Lang{qubo} into \Lang{maxsat} is
simpler. The following rule will transform a
\Lang{qubo} into a \Lang{max2sat} instance without increasing the
primal or incidence treewidth by more than one. Given a Hamiltonian
\[
H(x_1,\dots,x_n) = \sum_{i=1}^nw_ix_i + \sum_{i=1}^n\sum_{j=i+1}^n w_{ij}x_ix_j
\]
we first translate the \emph{minimization} problem of finding the
ground state into a \emph{maximization problem} by multiplying all
weights with $-1$. Then, we obtain a \Lang{wcnf} using:
\begin{reductionrule}\label{rule:qubo2maxsat}
  For every $x_i$ in $H$ we introduce a variable $x_i$, for every
  $i\in\{1,\dots,n\}$ and $j\in\{i+1,\dots,n\}$ we introduce an
  auxiliary variable $\alpha_{ij}$. Then we replace:
  \begin{itemize}
  \item every term $w_ix_i$ with $w_i>0$ by a clause $(x_i)^{w_i}$;
  \item every term $w_ix_i$ with $w_i<0$ by a clause $(\neg
    x_i)^{-w_i}$;
  \item every term $w_{ij}x_ix_j$ with $w_{ij}>0$ by clauses
    $(\alpha_{ij})^{w_{ij}}$, $(\alpha_{ij}\rightarrow x_i)^\infty$, and $(\alpha_{ij}\rightarrow x_j)^\infty$;
  \item every term $w_{ij}x_ix_j$ with $w_{ij}<0$ by clauses
    $(\neg\alpha_{ij})^{-w_{ij}}$, $(\neg\alpha_{ij}\rightarrow
    x_i)^\infty$, and $(\neg\alpha_{ij}\rightarrow x_j)^\infty$.
  \end{itemize}
\end{reductionrule}

\begin{lemma}\label{lemma:qubo2maxsat}
  Rule~\ref{rule:qubo2maxsat} produces a \Lang{wcnf} $\phi$ whose optimal
  models correspond to the ground states of $H$. The rule can be
  implemented in linear time and increases the primal and
  incidence treewidth by at most one.
\end{lemma}
\begin{proof}
  The runtime follows as for previous rules, as we simply replace
  terms by a constant number of new clauses.  
  As discussed before the rule, we can assume that the ground state of
  $H$ corresponds to a maximum point by multiplying all weights
  by $-1$. That the translation to \Lang{maxsat} is correct for all
  terms with $w>0$ is evident. The logic is inverted for the remaining
  cases: It is ``bad'' to satisfy the terms of~$H$, so it is ``good'' to
  satisfy the inverse. 
  
  Finally, it is easy to see that the primal treewidth increases
  by at most one, as for every term $w_{ij}x_ix_j$ there is a bag in the
  primal tree decomposition of $H$ that contains $x_i$ and $x_j$. By
  attaching a new bag containing $x_i$, $x_j$, and $\alpha_{ij}$, we
  obtain a primal tree decomposition of $\phi$.

  To see that the incidence treewidth is increased by at most one, take
  any tree decomposition of $I_\phi$.  All the cases of the rule are symmetric, so
  we only discuss the case in which the term $d=w_{ij}x_ix_j$ gets
  replaced by $(\alpha_{ij})^{w_{ij}}$, $(\alpha_{ij}\rightarrow
  x_i)^\infty$, and $(\alpha_{ij}\rightarrow x_j)^\infty$.
  We replace in every bag the term $d$ with $\alpha_{ij}$ and
  obtain two bags $t_1$ and $t_2$ with $\{\alpha_{ij},x_i\}\subseteq\bag(t_1)$ and
  $\{\alpha_{ij},x_j\}\subseteq\bag(t_2)$. To $t_1$ we attach two new
  bags, one containing $\alpha_{ij}$, $x_i$, and $(\alpha_{ij}\rightarrow
  x_i)^\infty$; and one that contains $\alpha_{ij}$ and $(\alpha_{ij})^{w_{ij}}$.
  Similarly, we attach a new bag to $t_2$ containing $\alpha_{ij}$, $x_j$, and $(\alpha_{ij}\rightarrow
  x_j)^\infty$.
\end{proof}

\begin{corollary}\label{cor:qubo}
  Rule~\ref{rule:qubo2maxsat} increases the primal treewidth to
  at most $\max(\tw,2)$ and the incidence treewidth to at most $\max(\itw,2)$.
\end{corollary}

\section{Intermezzo: Upper Bounds for Max2SAT}

Our reduction rules allow us to establish lower bounds for fragments
of \Lang{maxsat} under $\Class{SETH}$ and $\Class{ETH}$. The following
section considers upper bounds for fragments of \Lang{maxsat}.
Before we get there, let us first review the complexity of \Lang{maxsat}
with respect to the parameter treewidth.

It is straightforward (and folklore) to develop algorithms to solve
\Lang{maxsat} in time $O\big(2^{\tw(\phi)}\tw(\phi)^2|\phi|\big)$ and
$O\big(2^{2\itw(\phi)}\itw(\phi)^2|\phi|\big)$. It is known 
that the polynomial dependency on the treewidth can be removed by implementing
the underlying dynamic programs more carefully~\cite{BodlaenderBL13,BovaCMS15,CapelliM18}:

\begin{fact}\label{fact:maxsattw}
  \Lang{maxsat} can be solved in time $O\big(2^{\tw(\phi)}|\phi|\big)$
  and $O\big(2^{2\itw(\phi)}|\phi|\big)$.
\end{fact}

Note the gap between the primal and incidence treewidth, i.e., the
second running time contains a factor of two in the exponent.
This gap is \emph{not} present in binary formulas:

\begin{lemma}\label{lemma:max2sattw}
  \Lang{max2sat} can be solved in time $O\big(2^{\itw(\phi)}|\phi|\big)$.
\end{lemma}
\begin{proof}
  In the incidence graph of a \Lang{w2cnf}, all vertices corresponding
  to clauses have a maximum degree of~2. By the \emph{almost simplicial
  rule}, contracting such a vertex to one of its neighbors cannot
  increase the treewidth past~2~\cite{BodlaenderKEG01}. However,
  contracting all vertices corresponding to a clause to one of their
  neighbors yields exactly the primal graph, and, hence, we have
  $\tw(\phi)\leq\itw(\phi)$. The claim follows by
  Fact~\ref{fact:maxsattw}.
\end{proof}

\begin{corollary}
  \Lang{qubo} can be solved in time $O\big(2^{\itw(H)}|H|\big)$.
\end{corollary}

Lemma~\ref{lemma:max2sattw} and the corollary establish Theorem~\ref{theorem:binaryitw}.
The next section establishes similar results for fragments of
\Lang{maxsat} with arbitrary large clauses.

\section{Cost-optimal Reasoning via Model Counting}\label{section:prob-reasining}

In this section, we present stronger upper bounds for fragments of
\Lang{maxsat} by reducing these problems to the model counting problem
\Lang{\#sat}, i.e., to the task of counting the number of satisfying
assignments of a (unweighted) \Lang{cnf}. We establish
Theorem~\ref{thm:max2sat} in the form of three lemmas, one per item in
the statement. The variants are defined below. 
\begin{itemize}
\item \Lang{unary-maxsat}: Weights are given in
  unary and are, therefore, polynomially bounded by the input size.
\item \Lang{mult-maxsat}: Weights are powers of $2$, and we optimize for the product of weights instead of the sum.
\item \Lang{lex-maxsat}: The goal is to compute the lexicographically largest model, in which we first optimize for the largest number of variables with the largest weight, then for those with the second-largest weight, and so on.
\end{itemize}

To encode these problems into \Lang{\#sat}, it will be convenient to
assume that the weights are defined on 
\emph{variables} rather than  \emph{clauses}, i.e.,
$w\colon\vars(\phi)\rightarrow\mathbb{N}$. In this setting, the costs
of an assignment are $\cost(\beta)=\sum_{x\not\in\beta}w(x)$, and we
let $w(\beta)=\sum_{x\in\beta}w(x)$.
By Rule \ref{rule:unitsoft}, this is not a restriction and
does not increase the incidence treewidth.

\paragraph{Unary-Weighted MaxSAT.} For unary weights, the reduction
works in two steps. First, we present a structure-aware elimination of
unary weights. 
Then, we use the lemma below to translate \Lang{maxsat} to
\Lang{\#sat} with only a constant increase of the incidence treewidth.

\begin{lemma}\label{lem:elimunary}
There is a linear-time reduction from \Lang{maxsat} with weights given in
unary to \Lang{maxsat} 
with weights at most one such that the incidence treewidth is
increased by at most one.
\end{lemma}
\begin{proof}
  Let $\phi$ be a \Lang{wcnf} whose weight function $w$ is given in unary.
  For every variable~$x\in\vars(\phi)$ with~$w(x)>1$ we use $w(x)$ fresh weighted variables $w_i$
  with weight~$1$. 
  We construct $w(x)$ clauses of the form $c_i=(x \vee \neg w_i)$ for every~$i\in\{1,\dots, w(x)\}$.
  This ensures that we can only get the weight of~$x$ (via weighted variables $w_i$) in case $x$ is set to true.
  Observe that this increases the incidence treewidth by at most~$1$, as the fresh variables $w_i$
  do not need to occur in a common decomposition bag.
  Indeed, one can manipulate any tree decomposition of~$I_\phi$
  by finding a bag containing~$x$, attaching new bags containing
  $\{x,c_i\}$, to which another child with $\{c_i, w_i\}$ is added.
\end{proof}

\begin{lemma}\label{lem:weights}
There is a linear-time reduction from \Lang{maxsat} with weights at most one to \Lang{\#sat} 
such that the incidence treewidth is increased by at most one.
\end{lemma}
\begin{proof}  
  Observe that there can be at most~$2^{|\vars(\phi)|}$ (satisfying) assignments of~$\phi$.
In order to support this largest number, we require a base weight larger than $2^{|\vars(\phi)|}$.
We use the base weight $2^{|\vars(\phi)|+1}$ and eliminate $1$-weights by extending~$\phi$ 
to a new formula~$\phi'$.
Precisely, for every variable~$x$ with~$w(x)=1$, we use $|\vars(\phi)|+1$ auxiliary variables 
of the form~$w_i^x$ and add clauses $c_i^x=(x \vee w_i^x)$ for~$i\in\{1,\dots,|\vars(\phi)|\}$.

Observe that the construction of~$\phi'$ ensures that every satisfying assignment~$\beta$ of~$\phi'$ 
is duplicated $(2^{|\vars(\phi)|+1})^{w(\beta)} = 2^{w(\beta)\cdot(|\vars(\phi)|+1)}$ times.
Indeed, the number~$\#(\phi')$ of satisfying assignments of~$\phi'$ is of the 
form~$\sum_{\beta\models \phi} 2^{w(\beta)\cdot(|\vars(\phi)|+1)}$,
which makes it convenient to compute the maximal weight.
One can check whether the maximal solution
has weight at least $w$ if $\smash{\frac{\#(\phi')}{2^{w\cdot(|\vars(\phi)|+1)}}>0}$.
Note that this division works in polynomial time, using binary-encoded digits
and we only need a polynomial number of these queries.
To see that the incidence treewidth is increased by at most
one, take for every variable $x$ with $w(x)=1$ a bag containing $x$,
attach for every $i\in\{1,\dots,|\vars(\phi)|\}$ a new bag containing
$\{x,c_i^x\}$, and attach to this bag  $\{c_i^x,w_i^x\}$.
\end{proof}

Observe that both lemmas do \emph{not} increase the incidence
treewidth if the formula is not trivial, i.e., if it contains at least
one clause. In particular, applying both lemmas increases the
incidence treewidth by at most one.
With both lemmas at hand, we obtain the following result. 

\begin{lemma}[First Part of Theorem~\ref{thm:max2sat}]
\label{lemma:unary}
\Lang{unary-maxsat} can be solved in time $O(2^{\itw(\phi)})\poly(|\phi|)$.
\end{lemma}
\begin{proof}
Apply Lemmas~\ref{lem:elimunary} and~\ref{lem:weights}, and use the
algorithm by Slivovsky and Szeider~\cite{SlivovskySzeider20} to compute
$\#(\phi')$ in time $O(2^{\itw(\phi')})\poly(|\phi'|)$.
\end{proof}

\paragraph{Multiplicative MaxSAT.} 
With multiplicative weights, we follow a similar construction.  However, we need to 
utilize binary search to end up with a polynomial-time algorithm. We obtain the following new result.
\begin{lemma}[Second Part of Theorem~\ref{thm:max2sat}]
\label{lemma:mult}
\Lang{mult-maxsat} can be solved in time $O(2^{\itw(\phi)})\poly(|\phi|)$.
\end{lemma}
\begin{proof}
We perform a reduction similarly to Lemma~\ref{lem:weights} but
require a binary search to determine the largest weight. Again, we
construct a formula $\phi'$ by extending $\phi$ using the base weight
$2^{|\vars(\phi)|+1}$. For every $x\in\vars(\phi)$ with weight~$2^{w(x)}$, we use $w(x)\cdot(|\vars(\phi)|+1)$ auxiliary variables 
of the form~$w_i^x$ and clauses $c_i^x=(x \vee w_i^x)$ for every~$i\in\{1,\dots, w(x)\cdot(|\vars(\phi)|+1)\}$.
The construction ensures that satisfying assignments~$\beta$
are duplicated $(2^{|\vars(\phi)|+1})^{w(\beta)} = 2^{w(\beta)\cdot(|\vars(\phi)|+1)}$ times.
Indeed, the number~$\#(\phi')$ of satisfying assignments of~$\phi'$ is of the 
form~$\sum_{\beta\models \phi} 2^{w(\beta)\cdot(|\vars(\phi)|+1)}$,
which makes it convenient to use binary search to compute the maximal weight.
We compute $\#(\phi')$  as in Lemma~\ref{lemma:unary} and,
during the binary search, check whether the maximal solution
has weight at least $2^w$, namely if $\smash{\frac{\#(\phi')}{2^{w\cdot(|\vars(\phi)|+1)}}>0}$.
Note that $w$ is polynomial in the input size, i.e., the binary search only needs a polynomial number of divisions.
To conclude the proof, observe that the claim for the incidence
treewidth is analog to Lemma~\ref{lemma:unary}.
\end{proof}

\paragraph{Lexicographic MaxSAT.} This variant can be directly instantiated as \Lang{mult-maxsat}. 
\begin{lemma}[Third Part of Theorem~\ref{thm:max2sat}]
\label{lemma:lex}
\Lang{lex-maxsat} can be solved in time $O(2^{\itw(\phi)})\poly(|\phi|)$.
\end{lemma}
\begin{proof}
We translate \Lang{lex-maxsat} to \Lang{mult-maxsat} as follows: For
the smallest weight, we use $2^0$ as multiplicative weight, for the
second-smallest $2^{|\vars(\varphi)|+1}$, and so on, which leaves the
largest (say, $n$-th smallest) weight with
$2^{n(|\vars(\varphi)|+1)}$.  Note that weights might reoccur if they
did in the original instance.  We can determine the lexicographically
largest weight via binary search, as in Lemma~\ref{lemma:mult}.
\end{proof}

%
% --- Conclusion ---
%
\section{Conclusion and Future Research}\label{section:conclusion}

The recent rise of Ising machines created a demand for efficient
encodings into \Lang{max2sat} and \Lang{qubo}. Reductions from
\Lang{maxsat} to \Lang{max2sat} were previously considered to be of only
theoretical interest~\cite{AnsoteguiL21} and, surprisingly, have not been
studied so far with structural parameters like treewidth in mind. It
is surprising in the sense that solving a \Lang{qubo} using an
\Lang{ipu} requires an embedding from the problem into the hardware,
which may only be possible if the Hamiltonian has small
treewidth~\cite{wang2014ollivier}. Hence, if any part of the chain
from \Lang{maxsat} to \Lang{max2sat} to \Lang{qubo} increases the treewidth
non-negligible, using an \Lang{ipu} may not be feasible at all.

We showed that \Lang{maxsat} is equivalent to \Lang{qubo} under
\emph{treewidth-preserving} reductions. Our reductions are
tight for primal treewidth because they introduce, at most, a
constant \emph{additive} factor to the treewidth. As an immediate
consequence, we obtain an algorithm to solve \Lang{qubo}s in time
$2^{\tw(H)}\poly(|H|)$. These reductions also imply new lower
bounds under $\Class{SETH}$ for \Lang{max2sat} and \Lang{qubo}, which are
tight for primal treewidth. Table~\ref{table:results}
summarizes all the lower and upper bounds we have achieved.

\begin{table}[htbp]
  \caption{An overview of the lower (under $\Class{SETH}$) and upper bounds established in
    this article for various versions of \Lang{maxsat}. A line is
    highlighted if the bounds match. Note that all bounds are $\Class{ETH}$-tight.}
  \label{table:results}\label{tab:maxinc}
  \centering
  \begin{tabular}{rll}
    \toprule
    Problem & Lower Bound & Upper Bound \\
    \midrule    
    \multirow{2}{*}{\Lang{maxsat}}
    & \color{fg}\scriptsize$\Omega(2^{\tw(\phi)})\poly(|\phi|)$ & \color{fg}\scriptsize$O(2^{\tw(\phi)})\poly(|\phi|)$  \\ \cmidrule{2-3}
    & \scriptsize$\Omega(2^{\itw(\phi)})\poly(|\phi|)$ & \scriptsize$O(2^{2\itw(\phi)})\poly(|\phi|)$   \\ \cmidrule{1-3}
    \multirow{2}{*}{\Lang{max2sat}}
    & \color{fg}\scriptsize$\Omega(2^{\tw(\phi)})\poly(|\phi|)$ & \color{fg}\scriptsize$O(2^{\tw(\phi)})\poly(|\phi|)$  \\ \cmidrule{2-3}
    & \scriptsize$\Omega(2^{\itw(\phi)/3})\poly(|\phi|)$ & \scriptsize$O(2^{\itw(\phi)})\poly(|\phi|)$   \\ \cmidrule{1-3}
    \multirow{2}{*}{\Lang{mon-max2sat}}
    & \color{fg}\scriptsize$\Omega(2^{\tw(\phi)})\poly(|\phi|)$ & \color{fg}\scriptsize$O(2^{\tw(\phi)})\poly(|\phi|)$  \\ \cmidrule{2-3}
    & \scriptsize$\Omega(2^{\itw(\phi)/3})\poly(|\phi|)$ & \scriptsize$O(2^{\itw(\phi)})\poly(|\phi|)$   \\ \cmidrule{1-3}
    \multirow{2}{*}{\Lang{unary-maxsat}}
    & \color{fg}\scriptsize$\Omega(2^{\tw(\phi)})\poly(|\phi|)$ & \color{fg}\scriptsize$O(2^{\tw(\phi)})\poly(|\phi|)$  \\ \cmidrule{2-3}
    & \color{fg}\scriptsize$\Omega(2^{\itw(\phi)})\poly(|\phi|)$ & \color{fg}\scriptsize$O(2^{\itw(\phi)})\poly(|\phi|)$   \\ \cmidrule{1-3}
    \multirow{2}{*}{\Lang{mult-maxsat}}
    & \color{fg}\scriptsize$\Omega(2^{\tw(\phi)})\poly(|\phi|)$ & \color{fg}\scriptsize$O(2^{\tw(\phi)})\poly(|\phi|)$  \\ \cmidrule{2-3}
    & \color{fg}\scriptsize$\Omega(2^{\itw(\phi)})\poly(|\phi|)$ & \color{fg}\scriptsize$O(2^{\itw(\phi)})\poly(|\phi|)$   \\ \cmidrule{1-3}
    \multirow{2}{*}{\Lang{lex-maxsat}}
    & \color{fg}\scriptsize$\Omega(2^{\tw(\phi)})\poly(|\phi|)$ & \color{fg}\scriptsize$O(2^{\tw(\phi)})\poly(|\phi|)$  \\ \cmidrule{2-3}
    & \color{fg}\scriptsize$\Omega(2^{\itw(\phi)})\poly(|\phi|)$ & \color{fg}\scriptsize$O(2^{\itw(\phi)})\poly(|\phi|)$   \\ \cmidrule{1-3}
    \multirow{2}{*}{\Lang{qubo}}
    & \color{fg}\scriptsize$\Omega(2^{\tw(H)})\poly(|H|)$ & \color{fg}\scriptsize$O(2^{\tw(H)})\poly(|H|)$  \\ \cmidrule{2-3}
    & \scriptsize$\Omega(2^{\itw(H)/3})\poly(|H|)$ & \scriptsize$O(2^{\itw(H)})\poly(|H|)$   \\ \bottomrule
  \end{tabular}
\end{table}

For the incidence treewidth, our reductions only introduce a
\emph{multiplicative} factor of three. This implies a
$2^{3\itw(\phi)}\poly(|\phi|)$ algorithm for \Lang{maxsat}~--~while
reductions with an additive increase of the incidence treewidth would
imply a $\Class{SETH}$-optimal algorithm of running time
$2^{\itw(\phi)}\poly(|\phi|)$. The first immediate question is whether the result for incidence
treewidth can be improved to increase the incidence treewidth
by only an \emph{additive} factor:

\begin{openproblem}
  Is there a polynomial-time algorithm that, given a formula $\phi$ and a
  width-$k$ tree decomposition of~$I_\phi$, outputs a \Lang{w2cnf}
  $\psi$ and a width-$\big(k+O(1)\big)$ tree decomposition of $I_{\psi}$ such that $\cost(\phi)=\cost(\psi)$?
\end{openproblem}

The second downside of the reduction rules for incidence treewidth is
that they require a tree decomposition. Unfortunately, this is already
the case for the reduction to \Lang{max3sat}.

\begin{openproblem}
  Is there a polynomial-time algorithm that maps \Lang{wcnf}s~$\phi$
  to \Lang{w3cnf}s $\psi$ such that $\itw(\psi)\leq O(\itw(\phi))$ and $\cost(\phi)=\cost(\psi)$?
\end{openproblem}

From a practical perspective, the next natural step in this line of research is an experimental
evaluation of the proposed approach using different \Lang{ipu}
architectures and to compare these hardware accelerators to
state-of-the-art \Lang{maxsat} solvers. From the automated
reasoning perspective, it would be interesting to study
\Lang{maxsat} solvers that actively utilize binary clauses. There may
be benchmarks on which such tools have an advantage in conjunction
with the proposed reductions.

%
% Bibliography
%
%\clearpage
\bibliography{main}

\begin{thebibliography}{10}

\bibitem{AkopyanSCAAMIND15}
Filipp Akopyan, Jun Sawada, Andrew Cassidy, Rodrigo Alvarez{-}Icaza, John~V.
  Arthur, Paul Merolla, Nabil Imam, Yutaka~Y. Nakamura, Pallab Datta, Gi{-}Joon
  Nam, Brian Taba, Michael~P. Beakes, Bernard Brezzo, Jente~B. Kuang, Rajit
  Manohar, William~P. Risk, Bryan~L. Jackson, and Dharmendra~S. Modha.
\newblock {TrueNorth: Design and Tool Flow of a 65 mW 1 Million Neuron
  Programmable Neurosynaptic Chip}.
\newblock {\em {IEEE} Trans. Comput. Aided Des. Integr. Circuits Syst.},
  34(10):1537--1557, 2015.
\newblock \href {https://doi.org/10.1109/TCAD.2015.2474396}
  {\path{doi:10.1109/TCAD.2015.2474396}}.

\bibitem{AlHraishawiRC23}
Hayder Al{-}Hraishawi, Junaid ur~Rehman, and Symeon Chatzinotas.
\newblock {Quantum Optimization Algorithm for {LEO} Satellite Communications
  Based on Cell-Free Massive {MIMO}}.
\newblock In {\em {IEEE} International Conference on Communications, {ICC} 2023
  - Workshops, Rome, Italy, May 28 - June 1, 2023}, pages 1759--1764, 2023.
\newblock \href {https://doi.org/10.1109/ICCWORKSHOPS57953.2023.10283753}
  {\path{doi:10.1109/ICCWORKSHOPS57953.2023.10283753}}.

\bibitem{AlomEMWT17a}
Md.~Zahangir Alom, Brian~Van Essen, Adam~T. Moody, David~Peter Widemann, and
  Tarek~M. Taha.
\newblock {Quadratic Unconstrained Binary Optimization {(QUBO)} on Neuromorphic
  Computing System}.
\newblock In {\em 2017 International Joint Conference on Neural Networks,
  {IJCNN} 2017, Anchorage, AK, USA, May 14-19, 2017}, pages 3922--3929, 2017.
\newblock \href {https://doi.org/10.1109/IJCNN.2017.7966350}
  {\path{doi:10.1109/IJCNN.2017.7966350}}.

\bibitem{alves2024satellite}
Wallace Alves-Martins, Eva Lagunas, Nicolas Skatchkovsky, Flor de~Guadalupe
  Ortiz~Gomez, Geoffrey Eappen, Osvaldo Simeone, Bipin Rajendran, and Symeon
  Chatzinotas.
\newblock {Satellite Adaptive Onboard Beamforming Using Neuromorphic
  Processors}.
\newblock In {\em IEEE International Symposium on Personal, Indoor and Mobile
  Radio Communications}. IEEE, Washington, United States, 2024.

\bibitem{AnsoteguiL21}
Carlos Ans{\'{o}}tegui and Jordi Levy.
\newblock {Reducing {SAT} to Max2SAT}.
\newblock In {\em Proceedings of the Thirtieth International Joint Conference
  on Artificial Intelligence, {IJCAI} 2021, Virtual Event / Montreal, Canada,
  19-27 August 2021}, pages 1367--1373, 2021.
\newblock \href {https://doi.org/10.24963/IJCAI.2021/189}
  {\path{doi:10.24963/IJCAI.2021/189}}.

\bibitem{abs-2403-00182}
Carlos Ans{\'{o}}tegui and Jordi Levy.
\newblock {SAT, Gadgets, Max2XOR, and Quantum Annealers}.
\newblock {\em CoRR}, abs/2403.00182, 2024.
\newblock \href {https://arxiv.org/abs/2403.00182} {\path{arXiv:2403.00182}},
  \href {https://doi.org/10.48550/ARXIV.2403.00182}
  {\path{doi:10.48550/ARXIV.2403.00182}}.

\bibitem{BannachH24}
Max Bannach and Markus Hecher.
\newblock {Structure-Guided Cube-and-Conquer for MaxSAT}.
\newblock In {\em {NASA} Formal Methods - 16th International Symposium, {NFM}
  2024, Moffett Field, CA, USA, June 4-6, 2024, Proceedings}, pages 3--20,
  2024.
\newblock \href {https://doi.org/10.1007/978-3-031-60698-4_1}
  {\path{doi:10.1007/978-3-031-60698-4_1}}.

\bibitem{BianCMRSV17}
Zhengbing Bian, Fabi{\'{a}}n~A. Chudak, William~G. Macready, Aidan Roy, Roberto
  Sebastiani, and Stefano Varotti.
\newblock {Solving {SAT} and MaxSAT with a Quantum Annealer: Foundations and a
  Preliminary Report}.
\newblock In {\em Frontiers of Combining Systems - 11th International
  Symposium, FroCoS 2017, Bras{\'{\i}}lia, Brazil, September 27-29, 2017,
  Proceedings}, pages 153--171, 2017.
\newblock \href {https://doi.org/10.1007/978-3-319-66167-4_9}
  {\path{doi:10.1007/978-3-319-66167-4_9}}.

\bibitem{BodlaenderBL13}
Hans~L. Bodlaender, Paul~S. Bonsma, and Daniel Lokshtanov.
\newblock {The Fine Details of Fast Dynamic Programming over Tree
  Decompositions}.
\newblock In {\em Parameterized and Exact Computation - 8th International
  Symposium, {IPEC} 2013, Sophia Antipolis, France, September 4-6, 2013,
  Revised Selected Papers}, pages 41--53, 2013.
\newblock \href {https://doi.org/10.1007/978-3-319-03898-8_5}
  {\path{doi:10.1007/978-3-319-03898-8_5}}.

\bibitem{BodlaenderKEG01}
Hans~L. Bodlaender, Arie M. C.~A. Koster, Frank van~den Eijkhof, and Linda~C.
  van~der Gaag.
\newblock {Pre-processing for Triangulation of Probabilistic Networks}.
\newblock In {\em 17th Conference in Uncertainty in Artificial Intelligence
  (UAI 2001)}, pages 32--39, 2001.

\bibitem{BovaCMS15}
Simone Bova, Florent Capelli, Stefan Mengel, and Friedrich Slivovsky.
\newblock {On Compiling CNFs into Structured Deterministic DNNFs}.
\newblock In {\em Theory and Applications of Satisfiability Testing - {SAT}
  2015 - 18th International Conference, Austin, TX, USA, September 24-27, 2015,
  Proceedings}, pages 199--214, 2015.
\newblock \href {https://doi.org/10.1007/978-3-319-24318-4_15}
  {\path{doi:10.1007/978-3-319-24318-4_15}}.

\bibitem{CapelliM18}
Florent Capelli and Stefan Mengel.
\newblock {Tractable {QBF} by Knowledge Compilation}.
\newblock In {\em 36th International Symposium on Theoretical Aspects of
  Computer Science, {STACS} 2019, March 13-16, 2019, Berlin, Germany}, pages
  18:1--18:16, 2019.
\newblock \href {https://doi.org/10.4230/LIPICS.STACS.2019.18}
  {\path{doi:10.4230/LIPICS.STACS.2019.18}}.

\bibitem{chancellor2016direct}
Nicholas Chancellor, Stefan Zohren, Paul~A Warburton, Simon~C Benjamin, and
  Stephen Roberts.
\newblock {A Direct Mapping of Max k-SAT and High Order Parity Checks to a
  Chimera Graph}.
\newblock {\em Scientific reports}, 6(1):37107, 2016.

\bibitem{Codognet23}
Philippe Codognet.
\newblock {Encoding the At-Most-One Constraint for {QUBO} and Quantum
  Annealing: Experiments with the N-Queens Problem}.
\newblock In {\em Companion Proceedings of the Conference on Genetic and
  Evolutionary Computation, {GECCO} 2023, Companion Volume, Lisbon, Portugal,
  July 15-19, 2023}, pages 2195--2202, 2023.
\newblock \href {https://doi.org/10.1145/3583133.3596394}
  {\path{doi:10.1145/3583133.3596394}}.

\bibitem{codognet2024comparing}
Philippe Codognet.
\newblock {Comparing QUBO Models for Quantum Annealing: Integer Encodings for
  Permutation Problems}.
\newblock {\em International Transactions in Operational Research}, 2024.

\bibitem{CoffrinNB19}
Carleton Coffrin, Harsha Nagarajan, and Russell Bent.
\newblock {Evaluating Ising Processing Units with Integer Programming}.
\newblock In {\em Integration of Constraint Programming, Artificial
  Intelligence, and Operations Research - 16th International Conference,
  {CPAIOR} 2019, Thessaloniki, Greece, June 4-7, 2019, Proceedings}, pages
  163--181, 2019.
\newblock \href {https://doi.org/10.1007/978-3-030-19212-9_11}
  {\path{doi:10.1007/978-3-030-19212-9_11}}.

\bibitem{CyganFKLMPPS15}
Marek Cygan, Fedor~V. Fomin, Lukasz Kowalik, Daniel Lokshtanov, D{\'{a}}niel
  Marx, Marcin Pilipczuk, Michal Pilipczuk, and Saket Saurabh.
\newblock {\em Parameterized Algorithms}.
\newblock Springer, 2015.
\newblock \href {https://doi.org/10.1007/978-3-319-21275-3}
  {\path{doi:10.1007/978-3-319-21275-3}}.

\bibitem{DatePSP19}
Prasanna Date, Robert~M. Patton, Catherine~D. Schuman, and Thomas~E. Potok.
\newblock {Efficiently Embedding {QUBO} Problems on Adiabatic Quantum
  Computers}.
\newblock {\em Quantum Inf. Process.}, 18(4):117, 2019.
\newblock \href {https://doi.org/10.1007/S11128-019-2236-3}
  {\path{doi:10.1007/S11128-019-2236-3}}.

\bibitem{DaviesWOSGJPR21}
Mike Davies, Andreas Wild, Garrick Orchard, Yulia Sandamirskaya, Gabriel
  A.~Fonseca Guerra, Prasad Joshi, Philipp Plank, and Sumedh~R. Risbud.
\newblock {Advancing Neuromorphic Computing With Loihi: {A} Survey of Results
  and Outlook}.
\newblock {\em Proc. {IEEE}}, 109(5):911--934, 2021.
\newblock \href {https://doi.org/10.1109/JPROC.2021.3067593}
  {\path{doi:10.1109/JPROC.2021.3067593}}.

\bibitem{GareyJS76}
M.~R. Garey, David~S. Johnson, and Larry~J. Stockmeyer.
\newblock {Some Simplified NP-Complete Graph Problems}.
\newblock {\em Theor. Comput. Sci.}, 1(3):237--267, 1976.
\newblock \href {https://doi.org/10.1016/0304-3975(76)90059-1}
  {\path{doi:10.1016/0304-3975(76)90059-1}}.

\bibitem{guillaume2022deep}
Alexandre Guillaume, Edwin~Y Goh, Mark~D Johnston, Brian~D Wilson, Anita
  Ramanan, Frances Tibble, and Brad Lackey.
\newblock {Deep Space Network Scheduling Using Quantum Annealing}.
\newblock {\em IEEE Transactions on Quantum Engineering}, 3:1--13, 2022.

\bibitem{IgnatievMM19}
Alexey Ignatiev, Ant{\'{o}}nio Morgado, and Jo{\~{a}}o Marques{-}Silva.
\newblock {{RC2:} an Efficient MaxSAT Solver}.
\newblock {\em J. Satisf. Boolean Model. Comput.}, 11(1):53--64, 2019.
\newblock \href {https://doi.org/10.3233/SAT190116}
  {\path{doi:10.3233/SAT190116}}.

\bibitem{johnson2011quantum}
Mark~W Johnson, Mohammad~HS Amin, Suzanne Gildert, Trevor Lanting, Firas Hamze,
  Neil Dickson, Richard Harris, Andrew~J Berkley, Jan Johansson, Paul Bunyk,
  et~al.
\newblock {Quantum Annealing with Manufactured Spins}.
\newblock {\em Nature}, 473(7346):194--198, 2011.

\bibitem{KashimataYHT24}
Tomoya Kashimata, Masaya Yamasaki, Ryo Hidaka, and Kosuke Tatsumura.
\newblock {Efficient and Scalable Architecture for Multiple-Chip Implementation
  of Simulated Bifurcation Machines}.
\newblock {\em {IEEE} Access}, 12:36606--36621, 2024.
\newblock \href {https://doi.org/10.1109/ACCESS.2024.3374089}
  {\path{doi:10.1109/ACCESS.2024.3374089}}.

\bibitem{Krom67}
M.~R. Krom.
\newblock {The Decision Problem for a Class of First-Order Formulas in Which
  all Disjunctions are Binary}.
\newblock {\em Mathematical Logic Quarterly}, 13(1-2):15--20, 1967.
\newblock \href {https://doi.org/10.1002/malq.19670130104}
  {\path{doi:10.1002/malq.19670130104}}.

\bibitem{KrugerM20}
Tom Kr{\"{u}}ger and Wolfgang Mauerer.
\newblock {Quantum Annealing-Based Software Components: An Experimental Case
  Study with {SAT} Solving}.
\newblock In {\em {ICSE} '20: 42nd International Conference on Software
  Engineering, Workshops, Seoul, Republic of Korea, 27 June - 19 July, 2020},
  pages 445--450, 2020.
\newblock \href {https://doi.org/10.1145/3387940.3391472}
  {\path{doi:10.1145/3387940.3391472}}.

\bibitem{LampisMM18}
Michael Lampis, Stefan Mengel, and Valia Mitsou.
\newblock {{QBF} as an Alternative to Courcelle's Theorem}.
\newblock In {\em Theory and Applications of Satisfiability Testing - {SAT}
  2018 - 21st International Conference, {SAT} 2018, Held as Part of the
  Federated Logic Conference, FloC 2018, Oxford, UK, July 9-12, 2018,
  Proceedings}, pages 235--252, 2018.
\newblock \href {https://doi.org/10.1007/978-3-319-94144-8_15}
  {\path{doi:10.1007/978-3-319-94144-8_15}}.

\bibitem{LokshtanovMS11}
Daniel Lokshtanov, D{\'{a}}niel Marx, and Saket Saurabh.
\newblock {Lower bounds based on the Exponential Time Hypothesis}.
\newblock {\em Bull. {EATCS}}, 105:41--72, 2011.

\bibitem{Marques-SilvaAGL11}
Jo{\~{a}}o Marques{-}Silva, Josep Argelich, Ana Gra{\c{c}}a, and In{\^{e}}s
  Lynce.
\newblock {Boolean Lexicographic Optimization: Algorithms {\&} Applications}.
\newblock {\em Ann. Math. Artif. Intell.}, 62(3-4):317--343, 2011.
\newblock \href {https://doi.org/10.1007/S10472-011-9233-2}
  {\path{doi:10.1007/S10472-011-9233-2}}.

\bibitem{Marques-SilvaIM17}
Jo{\~{a}}o Marques{-}Silva, Alexey Ignatiev, and Ant{\'{o}}nio Morgado.
\newblock {Horn Maximum Satisfiability: Reductions, Algorithms and
  Applications}.
\newblock In {\em Progress in Artificial Intelligence - 18th {EPIA} Conference
  on Artificial Intelligence, {EPIA} 2017, Porto, Portugal, September 5-8,
  2017, Proceedings}, pages 681--694, 2017.
\newblock \href {https://doi.org/10.1007/978-3-319-65340-2}
  {\path{doi:10.1007/978-3-319-65340-2}}.

\bibitem{MartinsML14}
Ruben Martins, Vasco~M. Manquinho, and In{\^{e}}s Lynce.
\newblock {Open-WBO: {A} Modular MaxSAT Solver}.
\newblock In {\em Theory and Applications of Satisfiability Testing - {SAT}
  2014 - 17th International Conference, Held as Part of the Vienna Summer of
  Logic, {VSL} 2014, Vienna, Austria, July 14-17, 2014. Proceedings}, pages
  438--445, 2014.
\newblock \href {https://doi.org/10.1007/978-3-319-09284-3_33}
  {\path{doi:10.1007/978-3-319-09284-3_33}}.

\bibitem{MatsubaraTMSWTT20}
Satoshi Matsubara, Motomu Takatsu, Toshiyuki Miyazawa, Takayuki Shibasaki,
  Yasuhiro Watanabe, Kazuya Takemoto, and Hirotaka Tamura.
\newblock {Digital Annealer for High-Speed Solving of Combinatorial
  Optimization Problems and its Applications}.
\newblock In {\em 25th Asia and South Pacific Design Automation Conference,
  {ASP-DAC} 2020, Beijing, China, January 13-16, 2020}, pages 667--672, 2020.
\newblock \href {https://doi.org/10.1109/ASP-DAC47756.2020.9045100}
  {\path{doi:10.1109/ASP-DAC47756.2020.9045100}}.

\bibitem{Mniszewski19}
Susan~M. Mniszewski.
\newblock {Graph Partitioning as Quadratic Unconstrained Binary Optimization
  {(QUBO)} on Spiking Neuromorphic Hardware}.
\newblock In {\em Proceedings of the International Conference on Neuromorphic
  Systems, {ICONS} 2019, Knoxville, Tennessee, USA, July 23-25, 2019}, pages
  4:1--4:5, 2019.
\newblock \href {https://doi.org/10.1145/3354265.3354269}
  {\path{doi:10.1145/3354265.3354269}}.

\bibitem{MoraglioGS22}
Alberto Moraglio, Serban Georgescu, and Przemyslaw Sadowski.
\newblock {AutoQubo: Data-Driven Automatic {QUBO} Generation}.
\newblock In {\em {GECCO} '22: Genetic and Evolutionary Computation Conference,
  Companion Volume, Boston, Massachusetts, USA, July 9 - 13, 2022}, pages
  2232--2239, 2022.
\newblock \href {https://doi.org/10.1145/3520304.3533965}
  {\path{doi:10.1145/3520304.3533965}}.

\bibitem{MorseK23}
Gregory Morse and Tam{\'{a}}s Kozsik.
\newblock {On Optimal {QUBO} Encoding of Boolean Logic, (Max-)3-SAT and
  (Max-)k-SAT with Integer Programming}.
\newblock In {\em Proceedings of the 7th International Conference on
  Algorithms, Computing and Systems, {ICACS} 2023, Larissa, Greece, October
  19-21, 2023}, pages 145--153, 2023.
\newblock \href {https://doi.org/10.1145/3631908.3631929}
  {\path{doi:10.1145/3631908.3631929}}.

\bibitem{NaoukiIT23}
Jonathan Naoukin, Murat Isik, and Karn Tiwari.
\newblock {A Survey Examining Neuromorphic Architecture in Space and Challenges
  from Radiation}.
\newblock {\em CoRR}, abs/2311.15006, 2023.
\newblock \href {https://arxiv.org/abs/2311.15006} {\path{arXiv:2311.15006}},
  \href {https://doi.org/10.48550/ARXIV.2311.15006}
  {\path{doi:10.48550/ARXIV.2311.15006}}.

\bibitem{NussleinGLF22}
Jonas N{\"{u}}{\ss}lein, Thomas Gabor, Claudia Linnhoff{-}Popien, and Sebastian
  Feld.
\newblock {Algorithmic {QUBO} Formulations for k-SAT and Hamiltonian Cycles}.
\newblock In {\em {GECCO} '22: Genetic and Evolutionary Computation Conference,
  Companion Volume, Boston, Massachusetts, USA, July 9 - 13, 2022}, pages
  2240--2246, 2022.
\newblock \href {https://doi.org/10.1145/3520304.3533952}
  {\path{doi:10.1145/3520304.3533952}}.

\bibitem{NussleinZGLF23}
Jonas N{\"{u}}{\ss}lein, Sebastian Zielinski, Thomas Gabor, Claudia
  Linnhoff{-}Popien, and Sebastian Feld.
\newblock {Solving (Max) 3-SAT via Quadratic Unconstrained Binary
  Optimization}.
\newblock In {\em Computational Science - {ICCS} 2023 - 23rd International
  Conference, Prague, Czech Republic, July 3-5, 2023, Proceedings, Part {V}},
  pages 34--47, 2023.
\newblock \href {https://doi.org/10.1007/978-3-031-36030-5_3}
  {\path{doi:10.1007/978-3-031-36030-5_3}}.

\bibitem{ortiz2024energy}
Flor Ortiz, Nicolas Skatchkovsky, Eva Lagunas, Wallace~A Martins, Geoffrey
  Eappen, Saed Daoud, Osvaldo Simeone, Bipin Rajendran, and Symeon Chatzinotas.
\newblock {Energy-Efficient On-Board Radio Resource Management for Satellite
  Communications via Neuromorphic Computing}.
\newblock {\em IEEE Transactions on Machine Learning in Communications and
  Networking}, 2024.

\bibitem{Piotrow20}
Marek Piotr{\'{o}}w.
\newblock {UWrMaxSat: Efficient Solver for MaxSAT and Pseudo-Boolean Problems}.
\newblock In {\em 32nd {IEEE} International Conference on Tools with Artificial
  Intelligence, {ICTAI} 2020, Baltimore, MD, USA, November 9-11, 2020}, pages
  132--136, 2020.
\newblock \href {https://doi.org/10.1109/ICTAI50040.2020.00031}
  {\path{doi:10.1109/ICTAI50040.2020.00031}}.

\bibitem{RodriguezFarresBALC24}
Pol Rodr{\'{\i}}guez{-}Farr{\'{e}}s, Rocco Ballester, Carlos Ans{\'{o}}tegui,
  Jordi Levy, and Jes{\'{u}}s Cerquides.
\newblock {Implementing 3-SAT Gadgets for Quantum Annealers with Random
  Instances}.
\newblock In {\em Computational Science - {ICCS} 2024 - 24th International
  Conference, Malaga, Spain, July 2-4, 2024, Proceedings, Part {VI}}, pages
  277--291, 2024.
\newblock \href {https://doi.org/10.1007/978-3-031-63778-0_20}
  {\path{doi:10.1007/978-3-031-63778-0_20}}.

\bibitem{Santra2014}
Siddhartha Santra, Gregory Quiroz, Greg~Ver Steeg, and Daniel~A Lidar.
\newblock {Max 2-SAT with up to 108 Qubits}.
\newblock {\em New Journal of Physics}, 16(4):045006, apr 2014.
\newblock \href {https://doi.org/10.1088/1367-2630/16/4/045006}
  {\path{doi:10.1088/1367-2630/16/4/045006}}.

\bibitem{SchumanKPMDK22}
Catherine~D. Schuman, Shruti~R. Kulkarni, Maryam Parsa, J.~Parker Mitchell,
  Prasanna Date, and Bill Kay.
\newblock {Opportunities for Neuromorphic Computing Algorithms and
  Applications}.
\newblock {\em Nat. Comput. Sci.}, 2(1):10--19, 2022.
\newblock \href {https://doi.org/10.1038/S43588-021-00184-Y}
  {\path{doi:10.1038/S43588-021-00184-Y}}.

\bibitem{SlivovskySzeider20}
Friedrich Slivovsky and Stefan Szeider.
\newblock {A Faster Algorithm for Propositional Model Counting Parameterized by
  Incidence Treewidth}.
\newblock In Luca Pulina and Martina Seidl, editors, {\em Theory and
  Applications of Satisfiability Testing - {SAT} 2020 - 23rd International
  Conference, Alghero, Italy, July 3-10, 2020, Proceedings}, volume 12178 of
  {\em Lecture Notes in Computer Science}, pages 267--276. Springer, 2020.
\newblock \href {https://doi.org/10.1007/978-3-030-51825-7_19}
  {\path{doi:10.1007/978-3-030-51825-7_19}}.

\bibitem{TrevisanSSW00}
Luca Trevisan, Gregory~B. Sorkin, Madhu Sudan, and David~P. Williamson.
\newblock {Gadgets, Approximation, and Linear Programming}.
\newblock {\em {SIAM} J. Comput.}, 29(6):2074--2097, 2000.
\newblock \href {https://doi.org/10.1137/S0097539797328847}
  {\path{doi:10.1137/S0097539797328847}}.

\bibitem{VenkateshMNKD24}
Supreeth~Mysore Venkatesh, Antonio Macaluso, Marlon Nuske, Matthias Klusch, and
  Andreas Dengel.
\newblock {Quantum Annealing-Based Algorithm for Efficient Coalition Formation
  Among LEO Satellites}.
\newblock {\em CoRR}, abs/2408.06007, 2024.
\newblock \href {https://arxiv.org/abs/2408.06007} {\path{arXiv:2408.06007}},
  \href {https://doi.org/10.48550/arXiv.2408.06007}
  {\path{doi:10.48550/arXiv.2408.06007}}.

\bibitem{wang2014ollivier}
Chi Wang, Edmond Jonckheere, and Todd Brun.
\newblock {Ollivier-Ricci Curvature and Fast Approximation to Treewidth in
  Embeddability of QUBO Problems}.
\newblock In {\em 2014 6th International Symposium on Communications, Control
  and Signal Processing (ISCCSP)}, pages 598--601. IEEE, 2014.

\bibitem{ZielinskiBNLF24}
Sebastian Zielinski, Magdalena Benkard, Jonas N{\"{u}}{\ss}lein, Claudia
  Linnhoff{-}Popien, and Sebastian Feld.
\newblock {{SATQUBOLIB:} {A} Python Framework for Creating and Benchmarking
  (Max-)3SAT QUBOs}.
\newblock In {\em Innovations for Community Services - 24th International
  Conference, {I4CS} 2024, Maastricht, The Netherlands, June 12-14, 2024,
  Proceedings}, pages 48--66, 2024.
\newblock \href {https://doi.org/10.1007/978-3-031-60433-1_4}
  {\path{doi:10.1007/978-3-031-60433-1_4}}.

\bibitem{ZielinskiNSGLF23}
Sebastian Zielinski, Jonas N{\"{u}}{\ss}lein, Jonas Stein, Thomas Gabor,
  Claudia Linnhoff{-}Popien, and Sebastian Feld.
\newblock {Influence of Different 3SAT-to-QUBO Transformations on the Solution
  Quality of Quantum Annealing: {A} Benchmark Study}.
\newblock In {\em Companion Proceedings of the Conference on Genetic and
  Evolutionary Computation, {GECCO} 2023, Companion Volume, Lisbon, Portugal,
  July 15-19, 2023}, pages 2263--2271, 2023.
\newblock \href {https://doi.org/10.1145/3583133.3596330}
  {\path{doi:10.1145/3583133.3596330}}.

\end{thebibliography}
\end{document}